\documentstyle[12pt]{article}
\newtheorem{prop}{Proposition}[section]
\newtheorem{theorem}{Theorem}[section]

\begin{document}

\baselineskip=14pt
\begin{titlepage}
\vspace*{0.0cm}

\begin{center}
{\bf{\large  Synchronization of globally-coupled phase
oscillators:\\ singularities and scaling for general couplings}}
\end{center}
\vspace{0.25in}
\begin{center}
John David Crawford \\
Department of Physics and Astronomy\\
University of Pittsburgh\\
Pittsburgh, PA 15260
\end{center}
\begin{center}
K.T.R. Davies\\
Department of Physics\\
Duquesne University\\
Pittsburgh, Pennsylvania  15282-1503
\end{center}
\vspace{0.5cm}


\begin{quote}
\begin{abstract}
The onset of collective behavior in a population
of globally coupled oscillators with randomly distributed frequencies is
studied for phase dynamical models with arbitrary coupling; the effect of a
stochastic temporal variation in the frequencies is also included. The
Fokker-Planck equation for the coupled Langevin system is reduced to a kinetic
equation for the oscillator distribution function. Instabilities of the
phase-incoherent state are studied by center manifold reduction to the
amplitude dynamics of the unstable modes. Depending on the coupling, the
coefficients in the normal form can be singular in the limit of weak
instability when the diffusive effect of the
noise is neglected. A detailed analysis of these singularities to all orders in
the normal form expansion is presented. Physically, the singularities are
interpreted as predicting an altered
scaling of the entrained component near the onset of synchronization.
\end{abstract}
 \end{quote}

\begin{center}
keywords: oscillators, bifurcation, symmetry
\end{center}
\vfill
\begin{center}
January 9, 1997
\end{center}

\end{titlepage}
\tableofcontents
\baselineskip=24pt

\section{Introduction}
A phase dynamics description of $N$ limit cycle oscillators is feasible for
sufficiently weak interaction.\cite{erkop,kur,ashwin} The uncoupled system is
attracted to an N-torus and for weak coupling this attracting
torus will persist; the oscillator phases $(\theta_1(t),\ldots,\theta_N(t))$
are coordinates on the perturbed torus. In this paper we study such systems
where the interaction is mediated by a global coupling.\cite{kur,ashwin} More
precisely, we consider models of the form,
\begin{equation}
\dot{\theta}_i=\omega_i+\frac{K}{N}\sum^{N}_{j=1} f(\theta_j-\theta_i) +
\xi_i(t)\label{eq:gcoupled}
\end{equation}
with the frequencies $\omega_i$ of the uncoupled oscillators drawn from a
distribution $g(\omega)$ characteristic of the population or array. For both
physical and mathematical reasons, it is interesting to include in
(\ref{eq:gcoupled}) the effect of extrinsic white noise $\xi_i(t)$ perturbing
the deterministic phase dynamics; this term is defined by the ensemble
averages $<\xi_i(t)>=0$ and $<\xi_i(s)\,\xi_j(t)> = 2D\delta_{ij}\delta(s-t)$.
When the oscillators are uncoupled ($K=0$), then the noise and the intrinsic
variation in the
frequencies spreads the population in phase. For suitably chosen coupling
functions $f(\phi)$ as the strength of the coupling $K$ increases, the
instantaneous
frequencies $\dot{\theta}_i$ may become entrained and the phase disorder of the
uncoupled system gives way to an entrained state with partial or total phase
coherence.

The collective oscillations produced by such frequency-entrained states
is of interest in physics, chemistry, biology and rather recently
neuroscience.\cite{kur,win,win2,stro1,gray} The phase dynamics model
(\ref{eq:gcoupled}) was popularized especially by Kuramoto and his
collaborators for the coupling $f(\phi)=\sin\phi$ since this choice
allowed a mathematically tractable description of the onset of
entrainment.\cite{kur}
For a coupling $f(\phi)$ of arbitrary form, the model is a special case of a
general
normal form obtained by averaging.\cite{ashwin}
Very recently, the
averaged equations for a series array of weakly interacting
Josephson junctions has been shown to
have the form (\ref{eq:gcoupled}) with $f(\phi)=\sin\phi$ at leading order
in the coupling strength $K$.\cite{wcs} This result extended similar studies
that had treated the junctions as identical oscillators.\cite{ssw,ws}

Most research on phase oscillators with global couplings has been
motivated by their appeal as theoretical models of synchronization rather
than as realistic descriptions of specific experimental systems. In this
literature, until very recently, the preoccupation with the simple coupling
function considered by
Kuramoto has been nearly univeral; see \cite{stro1,kn} for discussions with
many references. Nevertheless, couplings $f(\phi)$ of more general form
are of interest for several reasons. First, interactions that are
derived when a reduction to phase dynamics is actually carried out can easily
have a more complicated structure.\cite{kur,hmm} Secondly, recent results
indicate that when the form of $f(\phi)$ is generalized, the emergence of the
entrained state can have qualitative different
features.\cite{dainew,jdc3,dainew96} More precisely, if $K_c$ is the
critical coupling, above which the synchronized state appears, then the
``size''
of the entrained component scales like $(K-K_c)^\beta$ above threshold.
In the absence of noise, Daido discovered $\beta=1$ when $f(\phi)$ contains
a second harmonic such as $\sin2\phi$, but for the Kuramoto model the
exponent drops to $\beta=1/2$, i.e. the entrainment for the more general
coupling can be much weaker at the same coupling stength $K$.\cite{dainew}
In this paper,  we present a detailed analysis of the first transistion to a
state with frequency entrainment that illuminates the interplay between
the intrinsic disorder $g(\omega)$, the external noise, and the coupling in
determining the exponent $\beta$. In particular, we make no further
restrictions
on $f(\phi)$ other than to assume it has a Fourier expansion
\begin{equation}
f(\phi)=\sum_{n=-\infty}^{\infty}\,f_n\,e^{in\phi}.\label{eq:fexp}
\end{equation}

For large oscillator populations, it is natural to analyze the system in the
limit $N\rightarrow\infty$ because of the relative simplicity afforded by
a continuum description of the population using a density function
$\rho(\theta,\omega,t)$. A convenient definition divides out the frequency
dependence of the uncoupled system so that
$g(\omega)\,\rho(\theta,\omega,t)\,d\theta\,d\omega$ describes
the fraction of oscillators with
natural frequency in $[\omega,\omega+d\omega]$
and phase in $[\theta,\theta+d\theta]$, and $\rho(\theta,\omega,t)\,d\theta$
describes
the fraction of oscillators at
natural frequency $\omega$
with phase in $[\theta,\theta+d\theta]$. This latter interpretation fixes the
normalization $\int\rho(\theta,\omega,t)\,d\theta=1$.
The dynamics
of the system (\ref{eq:gcoupled}) is formulated as a kinetic equation for
$\rho(\theta,\omega,t)$,
\begin{equation}
\frac{\partial\rho}{\partial t}+\frac{\partial(\rho v)}{\partial \theta}=D
\frac{\partial^2\rho}{\partial \theta^2},\label{eq:eveqn}
\end{equation}
with the deterministic part of the phase velocity
(\ref{eq:gcoupled}) expressed as an integral over the population
\begin{equation}
v(\theta,\omega,t)=\omega +K \int^{2\pi}_0\,d\theta'
\int^{\infty}_{-\infty}\,d\omega'
f(\theta'-\theta)\rho(\theta',\omega',t)\,g(\omega').\label{eq:vel}
\end{equation}
This equation has been applied recently to the Kuramoto
model;\cite{sak}-\cite{jdc0} our application to (\ref{eq:gcoupled}) for
arbitrary coupling is a
generalization of the results for $f(\phi)=\sin\phi$.\cite{sm,jdc0}
For completeness, the derivation of (\ref{eq:eveqn}) - (\ref{eq:vel}) is
outlined in Appendix \ref{app:deriv}.

The kinetic equation has a special solution,
\begin{equation}
\rho=\rho_0\equiv\frac{1}{2\pi},
\end{equation}
which describes a population of oscillators spread out unpreferentially in
phase. This uniform or incoherent distribution is an equilibrium since
$v(\theta,\omega)=\omega+Kf_0$ for $\rho=\rho_0$ implies $\partial_t\rho=0$.
The occurrence of instabilities of $\rho_0$ leading to
states with nontrivial phase dependence is established by a linear stability
analysis of $\rho_0$ in section \ref{sec:linear} and a nonlinear analysis of
the instabilities in sections \ref{sec:ampeqn} - \ref{sec:special}.
The nonlinear analysis focuses on the evolution of the linearly
unstable mode described by the flow on the unstable manifold of the
equilibrium; the amplitude equation describing this flow are formulated and
studied in
general terms in section \ref{sec:ampeqn}. The detailed calculation of the
unstable manifold flow given in section \ref{sec:exp} relies on expansions in
the amplitude of the unstable mode.  The exponent $\beta$ emerges as the
scaling exponent required to balance the linear instability against the
nonlinear terms in the amplitude equation. In the regime of weak linear
instability, different exponents can arise
because the behavior of the nonlinear terms varies considerably depending on
the form of the coupling and strength of the external noise. These features of
the expansion are
studied in sections \ref{sec:p1p2} - \ref{sec:special}.

Our conclusions regarding the properties of the amplitude equation are valid
to all orders in the expansion. This level of generality is possible in part
because the recursion relations for the expansion
are considerably simplified by the symmetry
of the problem. The model (\ref{eq:eveqn}) - (\ref{eq:vel}) has ${\rm SO}(2)$
or ${\rm O}(2)$ symmetry depending on details of the population and the
coupling.
The group ${\rm O}(2)$ is
generated by rotations $\beta\cdot(\theta,\omega)=(\theta +\beta,\omega)$
and reflections $\kappa\cdot(\theta,\omega)=-(\theta,\omega)$
which act on functions $\rho(\theta,\omega)$ in the usual way: for
any transformation $\chi\in{\rm O}(2)$,
$(\chi\cdot\rho)(\theta,\omega)\equiv\rho(\chi^{-1}\cdot(\theta,\omega))$. The
dynamics (\ref{eq:eveqn}) - (\ref{eq:vel}) is equivariant with respect to
rotations for
arbitrary choices of $g(\omega)$ and $f(\phi)$; if, in addition,
$g(\omega)=g(-\omega)$ {\em and} $f(\phi)=-f(-\phi)$, then the system commutes
with the reflection $\kappa$ as well. In the latter circumstance
the model has ${\rm O}(2)$ symmetry, otherwise the rotational
symmetry alone corresponds to ${{\rm SO}(2)}$. Note that the mean frequency
of $g(\omega)$ can always be shifted to zero and we assume that this has been
done.

The onset of linear instability for the incoherent state $\rho_0$ has some
unusual features, especially in the limit of weak noise $D\rightarrow0$.
This was first noticed in the context of the
Kuramoto model,\cite{sm}-\cite{jdc0} and is emphasized in the linear
analysis of section \ref{sec:linear}. In the absence of noise,
the unstable modes correspond to
eigenvalues emerging from a neutral continuum at onset; thus, the standard
framework of center manifold reduction cannot be directly applied. We
proceed by first deriving the amplitude equation on the center-unstable
manifold
with $D>0$, in this case the diffusion term (\ref{eq:eveqn}) shifts the
continuous
spectrum off the imaginary axis and the center manifold reduction is
straightforward.\cite{craw5,iv} The resulting amplitude equation is
then examined by first reducing the noise $D\rightarrow0$ and then
letting the linear growth rate of the mode approach zero
$\gamma\rightarrow0^+$.
In this way, we start with an amplitude equation on the unstable manifold and
examine its features as the unstable eigenvalues are allowed to approach a
neutral continuous spectrum on the imaginary axis.

Our procedure for extracting $\beta$ from the amplitude equation can be
illustrated by a simple example. Let $A(t)$ denote the mode amplitude
satisfying
\begin{equation}
\dot{A}(t)=\gamma\,A+a_1(\gamma)A^3+a_2(\gamma)A^5+\cdots\label{eq:aeqnintro}
\end{equation}
where the nonlinear coefficients $a_1, a_2,\ldots$ depend on the parameters of
the problem such as the growth rate $\gamma$. For small $\gamma$, this equation
should be viewed as a singular perturbation problem with the linear term
$\gamma\,A$ representing the perturbation. If $\gamma>0$, then there is always
a neighborhood of the equilibrium $A=0$ where the perturbation dominates the
unperturbed system and can completely change the dynamics.
As in other such singular problems, a possible strategy is to seek a (singular)
change of variables which transforms (\ref{eq:aeqnintro}) into a regular
perturbation problem. Thus it is natural to define a new amplitude
by
\begin{equation}
{A}(t)=\gamma^\beta r(\gamma^\delta t)\label{eq:newvar}
\end{equation}
and rewrite (\ref{eq:aeqnintro}) in terms of $r(\tau)$
\begin{equation}
\frac{d r}{d\tau} =\gamma^{1-\delta}\,r(\tau) + \gamma^{2\beta-\delta}
a_1(\gamma)r^3+\gamma^{4\beta-\delta}a_2(\gamma)r^5+\cdots.\label{eq:aeqnb}
\end{equation}
If possible, the choice of $\beta$ and $\delta$ should be made so that
each term in (\ref{eq:aeqnb}) is well behaved as $\gamma\rightarrow0^+$,
and furthermore so that the effect of $\gamma>0$ is a regular perturbation
of the system at $\gamma=0$.  The standard values are of course $\beta=1/2$
and $\delta=1$ yielding
\begin{equation}
\frac{d r}{d\tau} =r(\tau) + a_1(\gamma)r^3
+\gamma\,a_2(\gamma)r^5+\cdots.\label{eq:aeqnc}
\end{equation}
The unperturbed system $\dot{r}=r+a_1(0)r^3$ now includes the linear term
and near $r=0$ the effect of small $\gamma$ is unimportant. The change of
variables (\ref{eq:newvar}) is singular at $\gamma=0$ as expected, but
nevertheless we have determined the overall scaling $\beta=1/2$ for nonlinear
solutions
near the equilibrium in the regime of weak instability. In addition, since
terms in (\ref{eq:aeqnb}) at fifth order or higher are at least ${\cal
O}(\gamma)$, the unperturbed problem truncates to a simple balance between
the linear instability and the dominant nonlinear term. Note that since
$\gamma=0$
at $K=K_c$, then just above threshold $\gamma\sim(K-K_c)$, and the definition
in (\ref{eq:newvar}) implies $|A|\sim(K-K_c)^\beta$ for the scaling of the mode
ampltude near onset.

The preceeding discussion contains the unstated assumption that the
coefficients $a_1, a_2,\ldots$ have
well-defined {\em finite} limits  as $\gamma\rightarrow0$. In the phase
dynamical model
(\ref{eq:eveqn}) - (\ref{eq:vel}) this assumption is correct for any choice
of $f(\phi)$ and $g(\omega)$ as long as $D>0$. However if we set
$D=0$ before taking the limit $\gamma\rightarrow0^+$, then the nonlinear
coefficients are singular in general although the details depend on
$f(\phi)$. These singularities require a different choice for $\beta$
to accomplish the desiderata enumerated above. It turns out that when
one has a coupling with only a single Fourier component such as
in the Kuramoto model then the expansion coefficients are nonsingular
to all orders. Thus single component couplings are special and very
atypical. The most typical or generic behavior is
$\beta=1$ as was found by Daido in the examples he
considered.\cite{dainew,dainew96}
We further discuss
the relation between our approach and his theory in section~\ref{sec:daido}.

A key aspect of the singular behavior encountered in this
problem is the  emergence of the critical eigenvalues from a neutrally stable
continuous spectrum.  It has become clear recently that this characteristic is
shared by a diverse set of additional examples which include the onset of
linear instability in collisionless plasma\cite{jdc1}-\cite{jdcaj} and in fluid
shear flows\cite{case2}-\cite{briggs}, instabilities of solitary
waves\cite{pegowein1}-\cite{pegwein3}, and bifurcations in ``mean field''
descriptions of the dynamics of bubble clouds in fluids\cite{russo}.
Aside from the present work, only for the collisionless plasma models have the
resulting singularities in
the amplitude expansion been fully analyzed. Our proofs in this paper are
adapted from the techniques used in \cite{jdc2}.

\section{Linear analysis}\label{sec:linear}
Let $\eta\equiv\rho(\theta,\omega,t)-\rho_0$ denote the deviation from
$\rho_0$,
and rewrite the dynamics (\ref{eq:eveqn}) - (\ref{eq:vel}) to describe the
evolution of $\eta$:
\begin{equation}
\frac{\partial\eta}{\partial t}={\rm \cal L}\eta +{{\rm \cal
N}({\eta})}\label{eq:dyn}
\end{equation}
where the linear and nonlinear operators are defined by
\begin{eqnarray}
{\rm \cal L}\eta&=& D \frac{\partial^2\eta}{\partial \theta^2} -
(\omega+K\,f_0)\frac{\partial\eta}{\partial \theta} +
\frac{K}{2\pi}\int^{2\pi}_0\,d\theta'\int^{\infty}_{-\infty}\,d\omega'
f'(\theta'-\theta)\,\eta(\theta',\omega',t)\,g(\omega')
\label{eq:lop}\\
{{\rm \cal N}({\eta})}&=&
K\int^{2\pi}_0\,d\theta'\int^{\infty}_{-\infty}\,d\omega'
\,\eta(\theta',\omega',t)\,g(\omega')
\left[\rule{0in}{0.25in}\eta(\theta,\omega,t)\,f'(\theta'-\theta)\right.
\label{eq:nop}\\
&&\hspace{3.0in}-\left.
\frac{\partial\eta}{\partial\theta}(\theta,\omega,t)\,f(\theta'-\theta)
\right].\nonumber
\end{eqnarray}
In (\ref{eq:lop}) - (\ref{eq:nop}), $f'(\phi)\equiv df/d\phi$, and note that
the normalization of $\rho$ implies
\begin{equation}
\int^{2\pi}_0\,d\theta\,\eta(\theta,\omega,t)=0.\label{eq:nomean}
\end{equation}

The equilibrium $\rho_0$ is invariant under translations $\beta$ and
reflections $\kappa$; hence the dynamics for $\eta$ has the same symmetries as
the original system. This can be verified explicitly by noting that the
operators ${\rm \cal L}$ and ${\rm \cal N}$ commute with rotations for
arbitrary choices of $g(\omega)$ and $f(\phi)$, and also commute with $\kappa$
when $g(\omega)=g(-\omega)$ and $f(\phi)=-f(-\phi)$.

Since $\rho_0$ is independent of $\theta$, the eigenfunctions ${\rm \cal
L}\Psi=\lambda\Psi$ have the form
\begin{equation}
\Psi(\theta,\omega)=\psi(\omega)\,e^{in\theta},
\end{equation}
with $\psi(\omega)$ required to satisfy the eigenvalue equation
\begin{equation}
L_n\psi=\lambda\,\psi,\label{eq:evleqn}
\end{equation}
where
\begin{equation}
L_n\psi\equiv-in\left[(\omega+K\,f_0-inD)\psi+{K\,f_n^\ast}
\int^{\infty}_{-\infty}\,d\omega'\,g(\omega')\,\psi(\omega')\right].
\end{equation}
Because the $n=0$ component of $\eta$ is zero (\ref{eq:nomean}),
we can restrict our consideration of (\ref{eq:evleqn}) to $n\neq0$.

The spectrum of $L_n$ corresponds to values of $\lambda$ where
the resolvent $R_n(\lambda)\equiv(\lambda-L_n)^{-1}$ fails to be a bounded
operator. By solving the equation $(\lambda-L_n)\hat{\phi}=\phi$ for
$\hat{\phi}(\omega)$ we obtain the
resolvent explicitly, i.e. $\hat{\phi}=R_n(\lambda)\phi$ with
\begin{equation}
R_n(\lambda)\phi=\frac{-i/n}{\omega+K  f_0-i\lambda/n-inD}
\left[\phi(\omega)-\frac{K  f_n^\ast}{\Lambda_n(i\lambda/n)}
\int_{-\infty}^\infty\frac{d\omega'\;g(\omega')\phi(\omega')}
{\omega'+K  f_0-i\lambda/n-inD}\right]\label{eq:resolv}
\end{equation}
where, in the second term, $\Lambda_n(z)$ denotes
\begin{equation}
{\Lambda_{n}\,({z})}\equiv 1+{K\,f_{n}^\ast}
\int^{\infty}_{-\infty}\,d\omega\frac{g(\omega)}{\omega+Kf_0-z-inD}.
\label{eq:Lam}
\end{equation}
The details of a completely rigorous discussion of $R_n(\lambda)$ will depend
of course on the choice of a specific function space for the problem, e.g.
square integrable functions of $\omega$.
For our purposes,
it suffices to ignore these technicalities and discuss the spectrum of $L_n$
carefully though heuristically.

There are two ways $R_n(\lambda)\phi$ can be singular:
when $i\lambda/n$ hits a zero of the function $\Lambda_{n}$ and when the
denominator in (\ref{eq:resolv}) vanishes, i.e. when $\lambda=-in(\omega+K
f_0-inD)$. Since $-\infty<\omega<\infty$, this latter set of values coincides
with the line ${\rm Re}(\lambda)=-n^2D$, while the first set of values are
typically discrete $\lambda=-inz_j$ since $\Lambda_n(z_j)=0$ will have discrete
solutions $\{z_1, z_2, \ldots\}$. These two components generally correspond to
the essential spectrum and the point spectrum (eigenvalues), respectively.

This distinction is clarified by solving the eigenvalue equation directly. Let
$\lambda=-inz$, then (\ref{eq:evleqn}) reads
\begin{equation}
(\omega+K\,f_0-z-inD)\psi(\omega)=
-K\,f_n^\ast\,\int^{\infty}_{-\infty}\,d\omega'\,g(\omega')\,\psi(\omega').
\label{eq:evleqnb}
\end{equation}
The possible solutions for ${\rm Im}(z)\neq-nD$ and ${\rm Im}(z)=-nD$ are quite
different and we discuss them separately.

For ${\rm Im}(z)\neq-nD$ or ${\rm Re}(\lambda)\neq-n^2D$, we can divide by
$(\omega+K\,f_0-z-inD)$ and obtain the form of the eigenfunction
\begin{equation}
\psi(\omega)=
\frac{-K\,f_n^\ast\,\int\,d\omega'\,g(\omega')\,\psi(\omega')}
{\omega+K\,f_0-z-inD}\label{eq:form}
\end{equation}
in terms of the constant $\int\,d\omega'\,g\,\psi$. A nontrivial solution
in (\ref{eq:form}) is
found only if this constant is non-zero, in which case we can adopt the
normalization
\begin{equation}
\int^{\infty}_{-\infty}\,d\omega'\,g(\omega')\,\psi(\omega')=1.\label{eq:ncond}
\end{equation}
Together (\ref{eq:form}) and (\ref{eq:ncond}) determine an eigenfunction
\begin{equation}
\psi(\omega)=
\frac{-K\,f_n^\ast}
{\omega+K\,f_0-z-inD}\label{eq:efcnsoln}
\end{equation}
{\em provided} the normalization condition (\ref{eq:ncond}) for $\psi$ can be
satisfied
self-consistently. This requirement is precisely
$\Lambda_n(z)=\Lambda_n(i\lambda/n)=0$. Thus a root $\Lambda_n(z)=0$
implies the existence of an eigenvalue $\lambda=-inz$ with eigenfunction
(\ref{eq:efcnsoln}).

For all choices of $g(\omega)$ and $f(\phi)$, $\Lambda_n(z)$ satisfies the
identity
\begin{equation}
{\Lambda_{n}\,({z})}^\ast={\Lambda_{-n}\,({z}^\ast)};
\label{eq:Lam1}
\end{equation}
in addition, with reflection symmetry $f(\phi)=-f(-\phi)$ and
$g(\omega)=g(-\omega)$, a further identity holds
\begin{equation}
{\Lambda_{n}\,({z})}={\Lambda_{-n}\,(-{z})}.
\label{eq:Lam2}
\end{equation}
The first identity implies that complex eigenvalues must come in conjugate
pairs $(\lambda, \lambda^\ast)$ with conjugate eigenvectors
$\Psi(\theta,\omega)$ and
$\Psi(\theta,\omega)^\ast$, respectively. For a real eigenvalue $({\rm
Re}(z)=0)$, the
first identity reduces to the second (\ref{eq:Lam2}) and implies
that the eigenvalue will always
have (at least) two linearly independent eigenvectors, $\Psi$ and
$\Psi^\ast$. When we have a complex eigenvalue for a reflection symmetric
population, then (\ref{eq:Lam2}) holds and
again there will be two linearly independent eigenvectors
$\Psi(\theta,\omega)$ and
$\Psi(-\theta,-\omega)$. Without reflection symmetry, the typical case is a
complex eigenvalue, but with reflection symmetry both real and complex
eigenvalues can occur generically.

The actual occurrence of roots for $\Lambda_n(z)$ is a detailed question
depending on the frequency distribution $g$ and the coupling; examples have
been found for which $\Lambda_n(z)$ has no roots.\cite{sm}
In the complex $z$ plane, $\Lambda_n(z)$ is an analytic function for ${\rm
Im}(z)\neq-nD$
with a discontinuity across ${\rm Im}(z)=-nD$ whenever
$g({\rm Re}(z)-Kf_0)\neq0$. The discontinuity signals a branch cut for
$\Lambda_n(z)$ and the magnitude of the discontinuity is given by the boundary
values at $ {\rm Re}(z)=r$:
\begin{equation}
\lim_{\epsilon\rightarrow0^+}\Lambda_n(r-inD\pm i\epsilon)=
1+{K\,f_{n}^\ast}\left[{\rm PV}
\int^{\infty}_{-\infty}\,d\omega\frac{g(\omega)}{\omega+Kf_0-r}\right]
\pm i \pi\,K f_{n}^\ast\, g(r-Kf_0).
\label{eq:bvalue}
\end{equation}
As the parameters of the coupling or frequency distribution $g$ are varied, the
roots of $\Lambda_n(z)$ generally change. In particular, roots can appear or
disappear at the branch cut.

When $D=0$, the appearance of a root coincides with the onset of linear
instability and the boundary values (\ref{eq:bvalue}) provide
closed form expressions for
the coupling $K_c$ and frequency $\Omega_c=-{\rm Im}(\lambda)$ at criticality.
Let $z_0=(\Omega+i\gamma)/n$ denote the root with $\gamma>0$, and
$0=\Lambda_n((\Omega+i\gamma)/n)$. As $\gamma\rightarrow0^+$, we obtain
\begin{equation}
0=1+{K_c\,f_{n}^\ast}\left[{\rm PV}
\int^{\infty}_{-\infty}\,d\omega\frac{g(\omega)}{\omega-\Omega'}
\right]
+ i\pi\, K_c f_{n}^\ast\, g(\Omega').
\label{eq:criticality}
\end{equation}
where $\Omega'=\Omega_c/n-K_cf_0$.
The real and imaginary parts of this equation must vanish separately and this
implies
\begin{eqnarray}
\frac{-{\rm Im}(f_n)}{\pi\,|f_n|^2\,g(\Omega')}&=&K_c\\
&&\nonumber\\
\frac{\pi\,{\rm Re}(f_n)g(\Omega')}{{\rm
Im}(f_n)}&=&\int_0^{\infty}\frac{d\Delta}{\Delta}
\left[g(\Omega'+\Delta)-g(\Omega'-\Delta)\right].
\end{eqnarray}
The second equation determines $\Omega'$ and then the first equation determines
$K_c$; for $n=1$ and $f_0=0$ these expressions
reduce to the formulas derived
by Daido in his order function analysis.\cite{dai93}

For ${\rm Im}(z)=-nD$ or ${\rm Re}(\lambda)=-n^2D$, the eigenvalue equation
(\ref{eq:evleqnb})
has no nonsingular solutions. One can introduce distributions as solutions
following the closely related spectral theory for the Vlasov equation, but
we shall not require this development for our study (cf. \cite{ch} for
details).

In general, the spectrum of ${\cal L}$ has both eigenvalues and a
continuous component: for each
mode number $n=1,2,\ldots$, there is a line of continuous spectrum at
$\mbox{\rm Re}\,\lambda=-n^2D$; in addition ${\rm \cal L}$ has eigenvalues when
the function $\Lambda_{n}$
has roots. We are specifically interested in eigenvalues
that cross the imaginary axis, signifying instabilities of $\rho_0$. Depending
on the coupling, this can occur for any Fourier component $n=l$, provided there
is a root $\Lambda_{l}\,({z_0})=0$
such that ${\rm Im}(z_0)$ becomes positive. Henceforth $l$ and $z_0$ refer to
the eigenvalue $\lambda=-ilz_0$ determining the instability and the unstable
eigenfunction is
\begin{eqnarray}
\Psi(\theta,\omega)&=&\psi(\omega)\,e^{il\theta}\\
\psi(\omega)&=&\frac{-K\,f_l^\ast}
{\omega+K\,f_0-z_0-ilD}.\label{eq:efcn0}
\end{eqnarray}
Clearly the coupling must have a Fourier component $f_l\neq0$.

We further restrict the
problem with two additional assumptions. First, the root $z_0$ is simple, i.e.
$\Lambda'_{l}\,({z_0})\neq0$ where $\Lambda'_{l}\equiv d\Lambda_{l}/dz;$
secondly, the center subspace at criticality is two-dimensional, in particular,
all other roots of $\Lambda_{n}$ for any $n>0$ have negative imaginary parts
and remain bounded away from the real axis. The first assumption ensures that
there
are no generalized eigenvectors; the second restricts the analysis to
populations with rotation symmetry (SO(2)) and complex eigenvalues, or
populations with rotation and reflection symmetry (O(2)) and a real
eigenvalue.\cite{jdc0}
This restriction excludes one possible codimension-one bifurcation, a Hopf
bifurcation with O(2) symmetry, which involves a four-dimensional
center subspace; a discussion of $O(2)$ Hopf for the Kuramoto model has been
given elsewhere.\cite{jdc0}

The projection of $\eta$ along the critical mode (\ref{eq:efcn0}) requires an
inner product
\begin{equation}
(A,B)\equiv\int_{0}^{2\pi}d\theta\int_{-\infty}^{\infty}
d\omega\,A(\theta,\omega)^\ast\,B(\theta,\omega),\label{eq:ip}
\end{equation}
so that we can define an adjoint operator for ${\rm \cal L}$
\begin{equation}
({\rm \cal L}^\dagger\,A,B)=(A,{\rm \cal L}\,B).
\end{equation}
In terms of the Fourier expansion  $A=\sum_n\,A_n\exp (in\theta$), the inner
product becomes $(A,B)=2\pi\,\sum_n\,<A_n,B_n>$ using a convenient notation for
the $\omega$ integration
\begin{equation}
<A_n,B_n>\equiv\int_{-\infty}^{\infty}
d\omega\,A_n(\omega)^\ast\,B_n(\omega).\label{eq:ipw}
\end{equation}
A simple calculation shows that
\begin{equation}
{\rm \cal L}^\dagger\,A=\sum_{n=-\infty}^\infty\,
e^{in\theta}L_n^\dagger A_n(\omega),
\end{equation}
with
\begin{equation}
(L_n^\dagger A_n)(\omega)=in\left[(\omega+K\,f_0+inD)A_n(\omega)
+K\,f_n\,g(\omega)\int_{-\infty}^\infty\,d\omega'\,A_n(\omega')\right].
\end{equation}

The discussion of the spectrum of $L_n^\dagger$ parallels the analysis
for  $L_n$, and we simply state the results we shall require. The root
$\Lambda_{l}\,({z_0})=0$ implies an eigenfunction
${\rm \cal L}^\dagger\tilde{\Psi}=\lambda^\ast\tilde{\Psi}$ where
\begin{eqnarray}
\tilde{\Psi}(\theta,\omega)&=&\frac{1}{2\pi}
\tilde{\psi}(\omega)\,e^{il\theta}\\
\tilde{\psi}(\omega)&=&\frac{-g(\omega)}
{\Lambda'_{{l}}\,({z_0})^\ast(\omega+Kf_0-z_0^\ast+ilD)}.\label{eq:adjef}
\end{eqnarray}
This adjoint eigenfunction satisfies $(\tilde{\Psi},\Psi)=1$ and defines the
projection, $\eta\rightarrow(\tilde{\Psi},\eta)\;\Psi$, from $\eta$ onto the
$\Psi$ component of $\eta$. This projection is the only aspect of the adjoint
theory that we need.

Before developing the nonlinear theory, we illustrate our discussion with a
simple example of the instabilities we intend to analyze. Let the oscillator
population be described by a Lorentzian frequency distribution,
\begin{equation}
g(\omega)=\frac{\Delta}{\pi}\,\left[\frac{1}{\omega^2+\Delta^2}\right],
\label{eq:lorentz}
\end{equation}
then for ${\rm Im}(z)\neq-lD$, $\Lambda_{l}\,({z})$ is easily evaluated by
contour integration to obtain
\begin{equation}
\Lambda_{l}\,({z})=\left\{\begin{array}{cc}
\frac{z-K(f_0+f_l^\ast)+i(lD-\Delta)}{z-Kf_0+i(lD-\Delta)}&
{\rm Im}(z)<-lD\\
&\\
\frac{z-K(f_0+f_l^\ast)+i(lD+\Delta)}{z-Kf_0+i(lD+\Delta)}&
{\rm Im}(z)>-lD.\end{array}\right.
\label{eq:loroots}
\end{equation}
For ${\rm Im}(z)<-lD$, there is a root at $z_s=K(f_0+f_l^\ast)-i(lD-\Delta)$
when the coupling satisfies $K\,{\rm Im}(f_l)>\Delta.$ The  eigenvalues
$\lambda=-ilz_s$ have negative
real parts and fall to the left of the continuous spectrum at $\mbox{\rm
Re}\,\lambda=-l^2D$. An eigenvalue sitting to the right of the continuous
spectrum
requires a root with ${\rm Im}(z)>-lD$; for this example $\Lambda_{l}\,({z})$
has one such
root at $z_u=K(f_0+f_l^\ast)-i(lD+\Delta)$ when the coupling satisfies
$K\,{\rm Im}(f_l)<-\Delta$. This root yields the eigenvalue
\begin{equation}
\lambda=-ilz_u=\gamma-i\Omega,\label{eq:eigen}
\end{equation}
where
\begin{eqnarray}
\gamma&=&-l\left[ lD+\Delta+K\,{\rm Im}(f_l)\right]\\
\Omega&=&lK[{\rm Re}(f_l)+ f_0].
\end{eqnarray}
These modes are linearly stable for $lD+\Delta+K\,{\rm Im}(f_l)>0$ and become
linearly unstable for $K>K_c$ where $K_c=-(lD+\Delta)/{\rm Im}(f_l)$. Note,
that we find eigenvalues of $L_l$ only when  $|K\,{\rm Im}(f_l)|>\Delta$; in
the limit
$|K\,{\rm Im}(f_l)|\rightarrow\Delta$ these eigenvalues approach the continuous
spectrum and are ``absorbed'' as the underlying root $z_u$ or $z_s$ crosses the
branch cut onto a different Riemann sheet of $\Lambda_{l}\,({z})$.

The diffusion due to external noise simplifies the spectrum by displacing the
lines of continuous spectrum into the left half plane. With $D>0$ the
eigenvalues associated with instabilities first emerge from the continuum and
then subsequently reach the imaginary axis. Thus, the critical eigenvalues are
isolated on the imaginary axis, and one has a critical spectrum of standard
form for bifurcation theory. If we let $D\rightarrow0$, then the unstable
eigenvalues
emerge directly from a continuous spectrum on the imaginary axis. One of our
central concerns is the effect of this continuous spectrum for $D=0$ on the
$\gamma\rightarrow0^+$ limit of the vector field on the center-unstable
manifold. This is investigated below
by deriving this vector field first for $D>0$ and
then analyzing the effect of taking the limit $\gamma\rightarrow0^+$ when
$D=0$.

\section{Amplitude equation on the unstable manifold}\label{sec:ampeqn}
With diffusion included, the appearance of an unstable mode in the spectrum can
be analyzed by a straightforward application of center manfold
reduction.\cite{craw5,iv}  This enables us to derive approximately the vector
field on the center-unstable manifold of $\rho_0$, and study the resulting
bifurcation. This vector field will depend on the parameters of the problem and
we shall examine its features as
the growth rate of the unstable mode is progressively weakened, a limit which
we denote by $\gamma\rightarrow0^+$. The properties of this limit
are strongly dependent on the form of the coupling $f(\phi)$ and whether the
external noise $D>0$ is held fixed as $\gamma$ vanishes.

Center
manifold reduction yields an amplitude equation describing the
time-asymptotic behavior of the unstable mode.  The mode  amplitude
$\alpha(t)\equiv(\tilde{\Psi},\eta)$ and describes the projection of $\eta$
onto the unstable eigenspace (\ref{eq:efcn0}):
\begin{equation}
\eta(\theta,\omega,t)=[\alpha(t)\Psi(\theta,\omega)+ {\rm c.c.}] +
S(\theta,\omega,t)\label{eq:decomp}
\end{equation}
where $(\tilde{\Psi},S)=0.$ In terms of $\alpha$ and $S$, the
evolution equation (\ref{eq:dyn}) becomes
\begin{eqnarray}
\dot{\alpha}&=&\lambda\, \alpha +(\tilde{\Psi},{{\rm \cal N}({\eta})})
\label{eq:alph1}\\
\frac{\partial S}{\partial t}&=&{\rm \cal L} S+{{\rm \cal N}({\eta})}
-\left[(\tilde{\Psi},{{\rm \cal N}({\eta})})\,\Psi  + {\rm c.c.} \right].
\label{eq:S1}
\end{eqnarray}
These coupled equations are equivalent to (\ref{eq:dyn}); by
restricting them to solutions on the center-unstable manifold we obtain an
autonomous two-dimensional flow for $\alpha(t)$.

Near the equilibrium $\eta=0$, the center-unstable manifold is
described by a function $H$,
\begin{equation}
S(\theta,\omega)=H(\theta,\omega,\alpha,\alpha^\ast),\label{graph}
\end{equation}
that satisfies
$0=H(\theta,\omega,0,0)=\partial_\alpha\,H(\theta,\omega,0,0)
=\partial_{\alpha^\ast}\,H(\theta,\omega,0,0)$. Using this
representation of the manifold, the general solution in
(\ref{eq:decomp}) can be adapted for
solutions on the manifold,
\begin{equation}
\eta^u(\theta,\omega,t)=[\alpha(t)\Psi(\theta,\omega)+ {\rm c.c.}] +
H(\theta,\omega,\alpha(t),\alpha(t)^\ast);\label{eq:umfd}
\end{equation}
the time dependence of $\eta^u(\theta,\omega,t)$ is entirely determined by
the dynamics of $\alpha(t)$.
An autonomous equation for $\alpha(t)$ follows next by restricting the
general equation (\ref{eq:alph1}) to solutions of the special form
(\ref{eq:umfd}):
\begin{equation}
\dot{\alpha}=\lambda\, \alpha +(\tilde{\Psi},{{\rm \cal N}({\eta^u})}).
\label{eq:alph1cm}
\end{equation}
This is the amplitude equation we wish to derive and analyze.

Certain general features of the amplitude equation follow from the symmetry
properties of the evolution equation (\ref{eq:dyn}) which are, of course,
shared by its expression in (\ref{eq:alph1}) - (\ref{eq:S1}). The action of the
rotations $\beta$ and reflection $\kappa$ on the variables
$(\alpha,S(\theta,\omega))$
can be determined from their definitions in (\ref{eq:decomp}). The rotation
$\beta\cdot(\theta,\omega)\rightarrow(\theta+\beta,\omega)$ of the phase
acts by
\begin{equation}
\beta\cdot(\alpha,S(\theta,\omega))=
(\alpha \,e^{-il\beta},S(\theta-\beta,\omega)),\label{eq:trans}
\end{equation}
and
\begin{equation}
\kappa\cdot(\alpha,S(\theta,\omega))=(\alpha^\ast,S(-\theta,-\omega))
\label{eq:ref}
\end{equation}
for the reflection.

The right hand side of the amplitude equation
commutes with these transformations of $\alpha$ whenever they are symmetries
of (\ref{eq:alph1}) - (\ref{eq:S1}); in particular the amplitude equation
commutes with rotations (\ref{eq:trans}). There are useful standard results on
the expression of such equivariant vector fields. For example, a
two-dimensional vector field $\dot{\alpha}=V(\alpha,\alpha^\ast)$ that commutes
with $\alpha\rightarrow \alpha \,e^{-il\beta}$ can be written as
$V(\alpha,\alpha^\ast)=\alpha\,p(\sigma)$ where $\sigma\equiv|\alpha|^2$ and
$p(\sigma)$ is a smooth function.\cite{gss} Hence we know the amplitude
equation (\ref{eq:alph1cm}) has the form
\begin{equation}
\lambda\, \alpha +(\tilde{\Psi},{{\rm \cal N}({\eta^u})})=\alpha\,p(\sigma)
\label{eq:ampsym}
\end{equation}
although $p(\sigma)$ must be determined from the model.  In general,
$p(\sigma)$ is complex, but when the model has reflection
symmetry then $p(\sigma)$ must be real-valued.

\subsection{Analysis of $H(\theta,\omega,\alpha,\alpha^\ast)$}

The dynamical invariance of the center-unstable
manifold implies an equation for the
graph function $H$. Consistency between the time dependence of $S$
in (\ref{graph}) and the evolution of $S$ described by
(\ref{eq:alph1})-(\ref{eq:S1}) requires that
$H(\theta,\omega,\alpha(t),\alpha^\ast(t))$ satisfy
\begin{equation}
\left[\dot{\alpha}\,\frac{\partial H}{\partial {\alpha}}
 +\dot{\alpha}^\ast\,\frac{\partial H}{\partial {\alpha}^\ast}\right]
={\cal{L}} H+{\cal{N}}(\eta^u)-\left[(\tilde{\Psi},{\cal{N}}(\eta^u))\,\Psi +
(\tilde{\Psi},{\cal{N}}(\eta^u))^\ast\,\Psi^\ast\right]\label{Heqn}
\end{equation}
where $\dot{\alpha}$ on the left hand side refers to the amplitude equation
(\ref{eq:alph1cm}).

There are general constraints on $H$ due to the symmetries of (\ref{eq:alph1})
- (\ref{eq:S1}) because we expect the unstable manifold to be mapped onto
itself by a symmetry transformation. For a rotation $\beta$,
this symmetry invariance of the manifold implies
\begin{equation}
H(\theta-\beta,\omega,\alpha,\alpha^\ast)=
H(\theta,\omega,\alpha \,e^{-il\beta},\alpha^\ast \,e^{il\beta}),
\label{eq:Htrans}
\end{equation}
and if reflection symmetry holds then $H$ must satisfy
\begin{equation}
H(-\theta,-\omega,\alpha,\alpha^\ast)=
H(\theta,\omega,\alpha^\ast ,\alpha)
\label{eq:Href}
\end{equation}
as well.

These relations constrain the form of the Fourier expansion of H
\begin{equation}
H(\theta,\omega,\alpha,\alpha^\ast)=
\sum_{n=-\infty}^{\infty}H_n(\omega,\alpha,\alpha^\ast)\,e^{in\theta};
\label{eq:scm}
\end{equation}
applying (\ref{eq:Htrans}) to the components $H_n$ shows that the
component must vanish unless $n$ is an integer multiple of $l$. Note that
the $n=0$ component of $H$
is also identically zero because of the general property
$\int\,d\theta\,\eta=0$. When $n=ml>0$, then
$H_{ml}$ must have the form
\begin{eqnarray}
H_l(\omega,\alpha,\alpha^\ast)&=&\alpha\,\sigma h_{1}(\omega,\sigma)
\label{eq:Hl}\\
H_{ml}(\omega,\alpha,\alpha^\ast)&=&\alpha^m\,h_{m}(\omega,\sigma)
\hspace{0.5in}m>1.\label{eq:Hml}
\end{eqnarray}
The functions $h_{m}(\omega,\sigma)$ are unconstrained by the
rotations, but must satisfy $h_{m}(-\omega,\sigma)^\ast=h_{m}(\omega,\sigma)$
when reflection symmetry holds. In addition, the general relation
$(\tilde{\Psi},S)=0$ becomes $(\tilde{\Psi},H)=0$ for solutions on the unstable
manifold, and for the components of $H$ this implies
\begin{equation}
<\tilde{\psi},h_{1}>=0.\label{eq:horthog}
\end{equation}

The symmetry argument for (\ref{eq:Hl}) and (\ref{eq:Hml}) can be briefly
summarized.  Applying (\ref{eq:Htrans}) to the Fourier coefficients of $H$
leads to the identity
\begin{equation}
H_n(\omega,\alpha,\alpha^\ast)e^{-in\beta}=
H_n(\omega,\alpha e^{-il\beta},\alpha^\ast e^{il\beta}).
\label{eq:ftrans}
\end{equation}
For $\beta=2\pi/l$ this immediately implies that $H_n=0$ unless $\exp(-i2\pi
n/l)=1$, hence $n$ must be an integer multiple of $l$.
Recall that if a  function $F(\omega,\alpha,\alpha^\ast)$ is rotation invariant
under phase shifts $\alpha\rightarrow\alpha\exp-il\beta$,
then it  can be written
as $F(\omega,\alpha,\alpha^\ast)=g(\omega,\sigma)$ where the form of
$g(\omega,\sigma)$ is determined by $F$.\cite{gss}  In this case, the invariant
function $(\alpha^\ast)^mH_{ml}(\omega,\alpha,\alpha^\ast)=g(\omega,\sigma)$
must have the form $g(\omega,\sigma)=\sigma^mh(\omega,\sigma)$ to accomodate
the
factor $(\alpha^\ast)^m$. Dividing by $(\alpha^\ast)^m$ gives
$H_{ml}(\omega,\alpha,\alpha^\ast)=\alpha^m\,h(\omega,\sigma)$ as in
(\ref{eq:Hml}). The general property $\partial_\alpha H(\theta,\omega,0,0)=0$
requires that $h(\omega,\sigma)$ for $H_l$ have the special form
$h(\omega,\sigma)=\sigma h_1(\omega,\sigma)$ as
in (\ref{eq:Hl}).

The equations for $h_m(\omega,\sigma)$ follow by substituting
(\ref{eq:Hl}) - (\ref{eq:Hml}) and (\ref{eq:ampsym}) into the equation
for $H$ (\ref{Heqn}).
The results of this straightforward calculation for $m=1, 2$ and  $m>2$ are
summarized below. In writing these equations two further notations prove
convenient. First, define $\Gamma_m$ by
\begin{equation}
\Gamma_{m}(\sigma)\equiv
\int_{-\infty}^\infty\,d\omega\,g(\omega)h_{m}(\omega,\sigma),\label{eq:defG}
\end{equation}
and secondly let ${\rm P}_\perp$ denote the projection operator
\begin{equation}
({\rm P}_\perp\phi)(\omega)\equiv \phi(\omega)-<\tilde{\psi},\phi>\psi(\omega)
\label{eq:projop}
\end{equation}
which projects $\phi$ onto the orthogonal
complement of the eigenvector $\psi$.

The $n=l$ component ($m=1$) of (\ref{Heqn}) can be written
\begin{eqnarray}
[L_l-2p-p^\ast]h_1-(p+p^\ast)\sigma\frac{\partial h_1}{\partial\sigma}&=&
(2\pi iKl)\;{\rm P}_\perp\left\{\rule{0.0in}{0.25in}
f_l(1+\sigma\Gamma_1^\ast)h_2
+f_{2l}^\ast\Gamma_2(\psi^\ast+\sigma h_1^\ast)\right.\nonumber\\
&&\hspace{0.75in}+f_{2l}\,\sigma\,\Gamma_2^\ast \,h_3
+f_{3l}^\ast\sigma\Gamma_3 h_2^\ast\label{eq:m=1}\\
&&\hspace{0.0in}\left.
+\sum_{m=3}^\infty\sigma^{m-1}\left[f_{ml}\Gamma_m^\ast h_{m+1}
+f_{(m+1)l}^\ast\Gamma_{m+1} h_m^\ast\right]\right\}.\nonumber
\end{eqnarray}
The $n=2l$ component of (\ref{Heqn}) is
\begin{eqnarray}
[L_{2l}-2p]h_2-(p+p^\ast)\sigma\frac{\partial h_2}{\partial\sigma}&=&
(2\pi iK(2l))\;\left\{\rule{0.0in}{0.25in}
f_l^\ast[\psi+\sigma(h_1+\Gamma_1\psi)+\sigma^2\Gamma_1h_1]\right.
\label{eq:m=2}\\
&&\hspace{0.1in}+f_l\sigma(1+\sigma\Gamma_1^\ast)h_3
+f_{3l}^\ast\sigma[\Gamma_3\psi^\ast+\sigma\Gamma_3h_1^\ast]
\nonumber\\
&&\hspace{0.1in}\left.
+\sum_{m=2}^\infty\sigma^mf_{ml}\Gamma_m^\ast h_{m+2}
+\sum_{m=4}^\infty\sigma^{m-2}f_{ml}^\ast\Gamma_{m}h_{m-2}^\ast
\right\}.\nonumber
\end{eqnarray}
For $m\geq3$, the Fourier components of (\ref{Heqn}) are
\begin{eqnarray}
[L_{ml}-mp]h_m-(p+p^\ast)\sigma\frac{\partial h_m}{\partial\sigma}&=&
(2\pi iK(ml))\;\left\{\rule{0.0in}{0.25in}
f_l^\ast[h_{m-1}+\sigma \Gamma_1h_{m-1}]
+f_l\sigma(1+\sigma\Gamma_1^\ast)h_{m+1}\right.\nonumber\\
&&\hspace{1.1in}
+\sum_{n=2}^{m-2}f_{(m-n)l}^\ast\Gamma_{m-n}h_n\label{eq:m>2}\\
&&\hspace{0.1in}
+f_{(m-1)l}^\ast\Gamma_{m-1}[\psi+\sigma h_1]
+f_{(m+1)l}^\ast\sigma\Gamma_{m+1}[\psi^\ast+\sigma h_1^\ast]\nonumber\\
&&\hspace{0.1in}\left.
+\sum_{n=2}^{\infty}f_{nl}\sigma^n\Gamma_{n}^\ast h_{m+n}
+\sum_{n=2}^{\infty}f_{(m+n)l}^\ast\sigma^n\Gamma_{m+n}h_n^\ast\right\}.
\nonumber
\end{eqnarray}
Here and below, it is understood that sums are omitted when the lower value of
the summation index exceeds the upper value, e.g. for $m=3$ the first sum
in (\ref{eq:m>2}) is omitted.

The symmetry constraints imply that solutions on the unstable manifold
(\ref{eq:umfd}) have the general form
\begin{eqnarray}
\eta^u(\theta,\omega,t)&=&
\left[\rule{0.0in}{0.25in}\alpha(t)\Psi(\theta,\omega) +
\alpha(t)\,\sigma\,h_1(\omega,\sigma(t))\,e^{il\theta}\right.\label{eq:umfdb}\\
&&\hspace{1.0in}\left.+
\sum_{m=2}^\infty\alpha^m(t)\,h_m(\omega,\sigma(t))\,e^{iml\theta}\right]
+ {\rm c.c.}\nonumber
\end{eqnarray}
with $\alpha(t)$ satisfying the amplitude equation (\ref{eq:alph1cm}).
The set of equations (\ref{eq:ampsym}) and (\ref{eq:m=1}) - (\ref{eq:m>2})
determine the functions $p(\sigma)$ and $h_m(\omega,\sigma)$, however they
cannot be solved explicitly except using the amplitude expansions introduced
in the next section. What has been achieved to this point is to formulate
the problem entirely in terms of functions of a single {\em real} variable
$\sigma$. This simplification is essential to the analysis to follow.

\section{Expansions and recursion relations}\label{sec:exp}
Our conclusions regarding the nonlinear development of the instability are
based on amplitude expansions for $p$ and $h_m(\omega,\sigma)$, and the
corresponding
expansion of $\Gamma_m$ from (\ref{eq:defG}):
\begin{eqnarray}
p(\sigma)&=&\sum_{j=0}^\infty p_j\,\sigma^j\label{eq:pseries}\\
h_m(\omega,\sigma)&=&\sum_{j=0}^\infty h_{m,j}(\omega)\,\sigma^j\\
\Gamma_{m}(\sigma)&=&<g,h_{m}>=\sum_{j=0}^\infty
\Gamma_{m,j}\,\sigma^j\label{eq:defg}
\end{eqnarray}
where $\Gamma_{m,j}\equiv\int\,d\omega\,g\,h_{m,j}$.
The orthogonality condition (\ref{eq:horthog}) implies
\begin{equation}
<\tilde{\psi},h_{1,j}>=0\label{eq:orthog}
\end{equation}
at each order in the expansion of $h_1$.

The coefficients $p_j$ and $h_{m,j}(\omega)$ are determined by recursively
solving (\ref{eq:ampsym}) and (\ref{eq:m=1}) - (\ref{eq:m>2}).
Relations for $p_j$ follow from (\ref{eq:ampsym}) using the general
solution form in (\ref{eq:umfdb}) and the expansions (\ref{eq:pseries}) -
(\ref{eq:defg}). At each order in $\sigma^j$, $j\geq1$, we solve
(\ref{eq:ampsym}) for $p_j$ and find
\begin{eqnarray}
p_j&=&-(2\pi i K  l)\left\{
f_l\left[<\tilde{\psi},h_{2,j-1}>+\sum_{n=0}^{j-2}
\Gamma_{1,n}^\ast<\tilde{\psi},h_{2,j-n-2}>\right]\right.\label{eq:pj}\\
&&\hspace{0.25in}
+f_{2l}^\ast \left[\Gamma_{2,j-1}<\tilde{\psi},\psi^\ast>+\sum_{n=0}^{j-2}
\Gamma_{2,n}<\tilde{\psi},h_{1,j-n-2}^\ast>\right]
+f_{2l}\sum_{n=0}^{j-2}\Gamma_{2,n}^\ast<\tilde{\psi},h_{3,j-n-2}>\nonumber\\
&&\hspace{0.25in}\left.+\sum_{m=3}^{j+1}f_{ml}^\ast\sum_{n=0}^{j-m+1}
\Gamma_{m,n}<\tilde{\psi},h_{m-1,j-n-m+1}^\ast>
+\sum_{m=3}^{j}
f_{ml}\sum_{n=0}^{j-m}\Gamma_{m,n}^\ast<\tilde{\psi},h_{m+1,j-n-m}>
\right\}.\nonumber
\end{eqnarray}

The functions $h_{m,j}(\omega)$ are determined similarly by substituting
(\ref{eq:pseries}) - (\ref{eq:defg}) into the component equations
(\ref{eq:m=1}) - (\ref{eq:m>2}). At every order, $h_{m,j}$ is then
obtained by applying the resolvent operator (\ref{eq:resolv}) to
certain auxilliary functions  $I_{m,j}(\omega)$ constructed from
lower order terms, i.e.
\begin{equation}
h_{m,j}=-R_{ml}(\mu_{m,j})\,I_{m,j}(\omega)\label{eq:gen}
\end{equation}
where
\begin{equation}
\mu_{m,j}\equiv(m+j)\lambda+j\lambda^\ast+\delta_{m,1}(\lambda+\lambda^\ast),
\label{eq:arg}
\end{equation}
and the detailed expressions for $I_{m,j}$ are provided below.

The action of the resolvent is given in (\ref{eq:resolv})
\begin{equation}
R_{ml}(\mu_{m,j})\,I_{m,j}=\frac{-i/ml}{\omega+K  f_0-\nu_{m,j}-imlD}
\left[I_{m,j}(\omega)-\frac{K  f_{ml}^\ast}{\Lambda_{ml}(\nu_{m,j})}
\int_{-\infty}^\infty\frac{d\omega'\;g(\omega')I_{m,j}(\omega')}
{\omega'+K  f_0-\nu_{m,j}-imlD}\right]\label{eq:R}
\end{equation}
with $\nu_{m,j}\equiv i\mu_{m,j}/ml$. It is important to note from
(\ref{eq:arg}) that $\nu_{m,j}$ can be rewritten
\begin{equation}
\nu_{m,j}=z_0+\frac{2i(j+\delta_{m,1})\gamma}{ml};\label{eq:numj}
\end{equation}
hence for $m\geq1$, the poles $\omega=\nu_{m,j}-K\,f_0+imlD$ in (\ref{eq:R})
all sit in the upper half of the complex $\omega$ plane
along the line ${\rm Re}(\omega)={\rm Re}(z_0)-K\,f_0$.

The coefficients $p_j$ are determined by the integrals $\Gamma_{m,j}$,
$<\tilde{\psi},h_{m,j}>$, $<\tilde{\psi},h_{m,j}^\ast>$, and
$<\tilde{\psi},\psi^\ast>$. As we are interested in the form of these
coefficients near the onset of the instability ($\gamma\rightarrow0$) when the
noise is neglected (i.e. $D=0$), it is important to determine the behavior of
such integrals in this limit. In particular, we will need to estimate any
singular behavior exhibited by each type of integral.

In the remainder of the paper, our analysis of the amplitude expansion will be
facilitated by two identities.
{}From the definition of $\Gamma_{m,j}$ in (\ref{eq:defg}) and the
resolvent expression for $h_{m,j}(\omega)$ in (\ref{eq:gen}), the following
useful relations are obtained
\begin{eqnarray}
\Gamma_{m,j}&=&\frac{i/ml}{\Lambda_{ml}(\nu_{m,j})}
\int^\infty_{-\infty}\;d\omega
\frac{g(\omega)\,I_{m,j}(\omega)}{(\omega+Kf_0-\nu_{m,j}-imlD)}
\label{eq:identity1}\\
&&\nonumber\\
h_{m,j}(\omega)&=&\frac{iI_{m,j}(\omega)}{ml(\omega+Kf_0-\nu_{m,j}-imlD)}-
\frac{K f_{ml}^\ast\,\Gamma_{m,j}}{(\omega+Kf_0-\nu_{m,j}-imlD)}.
\label{eq:identity2}
\end{eqnarray}
Given $I_{m,j}$, these identities allow the properties of $h_{m,j}$ and
$\Gamma_{m,j}$ to be quickly determined.

The $I_{m,j}$ are in turn calculated from the following recursion relations
in terms of lower order coefficients. The application of these recursion
relations to calculate the coefficients $p_j$ is summarized in Table 1 and
illustrated in Section \ref{sec:p1p2}.

\subsection{Recursion relations for $I_{m,j}$}

The expansion of (\ref{eq:m=1}) using (\ref{eq:pseries}) - (\ref{eq:defg})
determines $h_{1,j}$ at each
order in (\ref{eq:R}): $h_{1,j}=-R_{l}(\mu_{1,j})\,I_{1,j}$ where
for $m=1$ and $j\geq0$,
\begin{eqnarray}
I_{1,j}&=&\sum_{k=0}^{j-1} [(2+k)p_{j-k}+(1+k)p_{j-k}^\ast]\,h_{1,k}
\label{eq:i1j}\\
&&\hspace{0.2in}+(2\pi i K  l)\;{\rm P}_\perp
\left\{f_{l}\left[h_{2,j}+\sum_{n=0}^{j-1}h_{2,n}\Gamma_{1,j-n-1}^\ast\right]
\right.\nonumber\\
&&\hspace{1.5in}+f_{2l}^\ast\left[\psi^\ast\Gamma_{2,j}+
\sum_{n=0}^{j-1}h_{1,n}^\ast\Gamma_{2,j-n-1}\right]
+f_{2l}\,\sum_{n=0}^{j-1}h_{3,n}\Gamma_{2,j-n-1}^\ast\nonumber\\
&&\hspace{0.5in}\left.
+\sum_{m=3}^{j+2}f_{ml}^\ast\,\sum_{n=0}^{j-m+2}\,
h_{m-1,n}^\ast\Gamma_{m,j-m-n+2}
+\sum_{m=3}^{j+1}f_{ml}\,\sum_{n=0}^{j-m+1}\,h_{m+1,n}\Gamma_{m,j-m-n+1}^\ast
\right\}.
\nonumber
\end{eqnarray}
Since $<\tilde{\psi},{\rm P}_\perp\phi>=0$ by construction, this recursion
relation together with the orthogonality property of $h_{1,j}$
(\ref{eq:orthog}) implies a corresponding orthogonality property for $I_{1,j}$
\begin{equation}
<\tilde{\psi},I_{1,j}>=0.\label{eq:Iorthog}
\end{equation}

For $m=2$, the expansion of (\ref{eq:m=2}) yields
$h_{2,j}=-R_{l}(\mu_{2,j})\,I_{2,j}$ where for $j=0$
\begin{eqnarray}
I_{2,0}&=&2\pi i K (2 l)\;f_l^\ast\;\psi(\omega),\label{eq:i20}
\end{eqnarray}
and for $j\geq1$,
\begin{eqnarray}
I_{2,j}&=&\sum_{k=0}^{j-1}\,[2p_{j-k}+k(p_{j-k}+p_{j-k}^\ast)]\,h_{2,k}
\label{eq:i2j}\\
&&\hspace{0.2in}+2\pi i K  (2l)\left\{f_l\left[h_{3,j-1}+
\sum_{n=0}^{j-2}\,h_{3,n}\,\Gamma_{1,j-n-2}^\ast\right]\right.\nonumber\\
&&\hspace{0.5in}+
f_l^\ast\left[h_{1,j-1}+\Gamma_{1,j-1}\,\psi
+\sum_{n=0}^{j-2}\,h_{1,n}\,\Gamma_{1,j-n-2}\right]
+f_{2l}\sum_{n=0}^{j-2}\,h_{4,n}\,\Gamma_{2,j-n-2}^\ast\nonumber\\
&&\hspace{0.5in}
+f_{3l}^\ast\left[\psi^\ast\Gamma_{3,j-1}
+\sum_{n=0}^{j-2}\,h_{1,n}^\ast\,\Gamma_{3,j-n-2}\right]
+f_{3l}\,\sum_{n=0}^{j-3}h_{5,n}\,\Gamma_{3,j-n-3}^\ast\nonumber\\
&&\hspace{0.5in}\left.
+\sum_{k=4}^{j+2}
f_{kl}^\ast\,\sum_{n=0}^{j-k+2}h_{k-2,n}^\ast\,\Gamma_{k,j-n-k+2}
+\sum_{k=4}^{j}
f_{kl}\,\sum_{n=0}^{j-k}h_{k+2,n}\,\Gamma_{k,j-n-k}^\ast\right\}.\nonumber
\end{eqnarray}

For $m\geq3$ and $j\geq0$ we have $h_{m,j}=-R_{ml}(\mu_{m,j})\,I_{m,j}$ with
\begin{eqnarray}
I_{m,j}&=&\sum_{k=0}^{j-1}\,[mp_{j-k}+k(p_{j-k}+p_{j-k}^\ast)]\,h_{m,k}
\label{eq:imj}\\
&&\hspace{0.2in}+2\pi i K (m l)
\left\{f_l\,\left[h_{m+1,j-1}+
\sum_{k=0}^{j-2}h_{m+1,k}\Gamma_{1,j-k-2}^\ast\right]
+f_l^\ast\,\left[h_{m-1,j}+
\sum_{k=0}^{j-1}h_{m-1,k}\Gamma_{1,j-k-1}\right]\right.\nonumber\\
&&\hspace{0.5in}+f_{(m-1)l}^\ast\,
\left[\psi\Gamma_{m-1,j}+\sum_{k=0}^{j-1}h_{1,k}\Gamma_{m-1,j-k-1}\right]
+\sum_{n=2}^{m-2}f_{nl}^\ast
\sum_{k=0}^{j}h_{m-n,k}\Gamma_{n,j-k}
\nonumber\\
&&\hspace{1.0in}
+f_{(m+1)l}^\ast
\left[\psi^\ast\Gamma_{m+1,j-1}+\sum_{k=0}^{j-2}h_{1,k}^\ast\Gamma_{m+1,j-k-2}
\right]\nonumber\\
&&\hspace{1.5in}\left.
+\sum_{n=2}^{j}f_{nl}\sum_{k=0}^{j-n}
h_{m+n,k}\Gamma_{n,j-n-k}^\ast
+\sum_{n=m+2}^{m+j}f_{nl}^\ast \sum_{k=0}^{j-n+m}
h_{n-m,k}^\ast\Gamma_{n,j-n-k+m}\right\}.\nonumber
\end{eqnarray}
In this final expression it is understood that a term with a negative subscript
is omitted, e.g. the term $f_l\,h_{m+1,j-1}$ is dropped when $j=0$.

\section{Pinching singularities at low order}\label{sec:p1p2}
At this point it is instructive to evaluate the low order coefficients in the
expansion and understand how the singularities arise mathematically. We first
calculate the cubic coefficient $p_1$ and show that it has a singularity
when $f_{2l}\neq0$. Then we evaluate the fifth order coefficient for the
special case of a coupling with $f_{2l}=0$ and no singularity in $p_1$.

\subsection{Evaluation of the coefficient $p_1$}\label{sec:p1}

The calculation of $p_{1}$ in (\ref{eq:pj}) yields
\begin{equation}
p_1=-2\pi iKl\left[f_l<\tilde{\psi},h_{2,0}>+
f_{2l}^\ast\Gamma_{2,0}<\tilde{\psi},\psi^\ast>\right]\label{eq:nfcoeff}
\end{equation}
where the function $h_{2,0}$ is determined from (\ref{eq:identity2}) and
(\ref{eq:i20})
\begin{equation}
h_{2,0}(\omega)=\frac{i\,I_{2,0}}{2l(\omega+Kf_0-z_0-i2lD)}
-\frac{K\,f_{2l}^\ast\,\Gamma_{2,0}}{(\omega+Kf_0-z_0-i2lD)}
\label{eq:cmcoeff}
\end{equation}
with $\Gamma_{2,0}$ given by (\ref{eq:identity1})
\begin{equation}
\Gamma_{2,0}=\frac{i/2l}{\Lambda_{{2l}}\,({z_0})}
\int\,d\omega\frac{g(\omega)\,I_{2,0}(\omega)}{(\omega+Kf_0-z_0-i2lD)}
=\frac{2\pi\,K\,f_{l}^\ast\,\Lambda'_{{l}}\,({z_0})}{\Lambda_{{2l}}\,({z_0})}
+{\cal O}(D).\label{eq:g20}
\end{equation}
The asymptotic expression for small $D$ in
(\ref{eq:g20}) shows that generically $\Gamma_{2,0}\neq0$.

The most important qualitative feature of $I_{2,0}(\omega)$ and
$h_{2,0}(\omega)$ is that they are meromorphic functions on the
complex $\omega$ plane. Let $z_0=(\Omega+i\gamma)/l$ denote the real and
imaginary parts of the root, then each function has a pole at
$\omega_+=(\Omega/l-Kf_0)+i(l^2D+\gamma)/l$. In addition,
$h_{2,0}(\omega)$ has a second pole at $(\Omega/l-Kf_0)+i(2l^2D+\gamma)/l$.
Both poles lie in the upper half plane and although they can approach the real
axis as the $\gamma\rightarrow0^+$, the integrals $<\tilde{\psi},h_{2,0}>$
and $\Gamma_{2,0}$ in (\ref{eq:nfcoeff}) have well-behaved nonsingular
limits even if $D$ is zero.

This observation regarding analytic structure can be extended considerably:
{\em all} of the functions $I_{m,j}(\omega)$ and $h_{m,j}(\omega)$ are
meromorphic functions in the complex $\omega$ plane and all of their poles lie
along the vertical line ${\rm Re}(\omega)=-Kf_0+{\rm Re}(z_0)$.  When $D=0$,
all of the poles approach the real axis in the limit
$\gamma\rightarrow0^+$;
in general, however the poles will be found in both the upper and lower half
plane and integrals over $I_{m,j}(\omega)$ or $h_{m,j}(\omega)$ will
exhibit pinching singularities. These conclusions follow from the recursion
relations in light of two simple observations. First, each $I_{m,j}(\omega)$ is
built up as a sum of various $h_{m,j}(\omega)$ from lower order and the sum
of two meromorphic functions is again meromorphic. Secondly, the application
of the resolvent (\ref{eq:R}) to a meromorphic $I_{m,j}(\omega)$ yields a
meromorphic function for $h_{m,j}(\omega)$ in (\ref{eq:identity2}).

The simplest pinching singularity is illustrated by $<\tilde{\psi},\psi^\ast>$
in $p_1$; from
(\ref{eq:efcn0}) and (\ref{eq:adjef}) one sees that
the integrand has poles $\omega_\pm=(\Omega/l-Kf_0)\pm i(l^2D+\gamma)/l$ above
and below the contour along the real axis. Now the $\gamma\rightarrow0^+$ limit
produces a pinching singularity when $D$ is zero, and this pinch
contributes a factor $(\gamma+l^2D)$ to the denominator when the integral is
evaluated:
\begin{equation}
<\tilde{\psi},\psi^\ast>=
-\frac{lf_l\,{\rm Im}(f_l)}{(\gamma+l^2D)\,|f_l|^2\,\Lambda'_{{l}}\,({z_0})}.
\label{eq:sing}
\end{equation}
This singularity will occur in the $p_1$ coefficient (\ref{eq:nfcoeff})
provided $D\approx0$ and $f_{2l}\neq0$.

The derivation of (\ref{eq:sing}) is simple and illustrates the key idea used
below in the general analysis: make a partial fraction expansion of the
integrand to isolate the singularity. In this case we have
\begin{equation}
<\tilde{\psi},\psi^\ast>=\frac{K\,f_l}{\Lambda'_{{l}}\,({z_0})}
\int^\infty_{-\infty}\,d\omega\,
\frac{g(\omega)}{(\omega-\omega_+)(\omega-\omega_-)}.
\label{eq:p1int}
\end{equation}
The partial fraction expansion yields
\begin{equation}
\frac{1}{(\omega-\omega_+)(\omega-\omega_-)}=
\frac{-il/2}{\gamma+l^2D}
\left(\frac{1}{(\omega-\omega_+)}-\frac{1}{(\omega-\omega_-)}\right),
\label{eq:pfex}
\end{equation}
so that (\ref{eq:p1int}) becomes
\begin{equation}
<\tilde{\psi},\psi^\ast>=\frac{-ilK\,f_l}{2\Lambda'_{{l}}({z_0})(\gamma+l^2D)}
\left(\frac{\Lambda_{{l}}({z_0})-1}{K\,f_l^\ast}-
\frac{\Lambda_{{l}}({z_0})^\ast-1}{K\,f_l}\right).
\label{eq:p1intb}
\end{equation}
This reduces to (\ref{eq:sing}) since $\Lambda_{{l}}({z_0})=0$.
For $D=0$, the cubic coefficient diverges like $p_1\sim\gamma^{-1}$ as
$\gamma\rightarrow0^+$ provided $f_{2l}\neq0$; this turns out to reflect a
general pattern $p_j\sim\gamma^{-2j+1}$ which is proved below in Theorem
\ref{thm:main}.

The significance of this divergence can be appreciated by considering the
amplitude equation truncated at third order
\begin{equation}
\dot{\alpha}=\lambda\alpha+p_1\,\alpha|\alpha|^2,
\end{equation}
and re-examining the determination of $\beta$ in (\ref{eq:aeqnintro}).
Introducing polar variables $\alpha=A(t)\exp(-i\psi(t))$, the amplitude
equation
reads
\begin{eqnarray}
\dot{A}&=&\gamma A+{\rm Re}(p_1)A^3\hspace{1.0in}
\dot{\psi}=\Omega -{\rm Im}(p_1)A^2\label{eq:polar}
\end{eqnarray}
where $\lambda=\gamma-i\Omega.$ Let $p_1=a_1\gamma^{-\nu}$ denote the
asymptotic
behavior of the cubic coefficient with $\nu\geq0$, then in terms of the
rescaled amplitude (\ref{eq:newvar}) the equations become
\begin{eqnarray}
\frac{d r}{d\tau}&=&\gamma^{1-\delta} r+\gamma^{2\beta-\nu-\delta}{\rm
Re}(a_1)r^3\label{eq:radial}
\hspace{1.0in}
\dot{\psi}=\Omega -\gamma^{2\beta-\nu}{\rm Im}(a_1)r^2.
\end{eqnarray}
In order to remove the singular perturbation due to the linear term we again
take $\delta=1$ and then solve for $\beta=(1+\nu)/2$ to balance the
cubic nonlinearity against the linear instability. This makes it clear that
a divergence in the cubic coefficient will increase $\beta$ and for $\nu=1$
we predict $\beta=1$. This of course neglects possible effects from
higher order nonlinear terms, in particular singularities at higher order
could turn out to be more important than the cubic singularity.  The effects of
higher order singularities are systematically studied in section
\ref{sec:sing}.

\subsection{The fifth order coefficient $p_2$}\label{sec:p2}

When $f_{2l}=0$ the cubic coefficient is nonsingular, and we briefly consider
what occurs at fifth order in the amplitude expansion. Setting $f_{2l}=0$, the
fifth order coefficient is given by
\begin{eqnarray}
p_2&=&-(2\pi i K  l)\left\{f_l\left[<\tilde{\psi},h_{2,1}>+
\Gamma_{1,0}^\ast<\tilde{\psi},h_{2,0}>\right] +
f_{3l}^\ast\Gamma_{3,0}<\tilde{\psi},h_{2,0}^\ast>\right\}.\label{eq:p2p}
\end{eqnarray}
where
\begin{eqnarray}
I_{1,0}(\omega)&=&(2\pi i K  l)\;{\rm P}_\perp
\left\{f_{l}h_{2,0}\right\}=(2\pi i K  l)\,f_{l}\;
\left\{h_{2,0}(\omega)-\psi(\omega)<\tilde{\psi},h_{2,0}>\right\}
\label{eq:I10}\\
I_{3,0}(\omega)&=&2\pi i K  (3l)\,
\;f_l^\ast\; h_{2,0}(\omega)\label{eq:I30}\\
I_{2,1}(\omega)&=&2p_1h_{2,0}(\omega)+2\pi i K  (2l)\,
\left[f_lh_{3,0}(\omega)+f_l^\ast(h_{1,0}(\omega)+\Gamma_{1,0}\psi(\omega))
\right.\\
&&\hspace{2.5in}\left.+
f_{3l}^\ast\Gamma_{3,0}\psi^\ast(\omega)\right].\nonumber
\end{eqnarray}

There are five integrals to consider in (\ref{eq:p2p}). Since $h_{2,0}$ has
poles only in the upper plane, the integral $<\tilde{\psi},h_{2,0}>$ is not
singular.
{}From (\ref{eq:I30}), we note that
$I_{3,0}$ similarly has poles only in the upper half plane hence $\Gamma_{3,0}$
has no pinching singularity. In addition, $I_{1,0}$ in (\ref{eq:I10})
has poles in
the upper plane only and this implies that  $\Gamma_{1,0}$ is nonsingular,
but the calculation is more subtle. From (\ref{eq:identity1}) we have
\begin{equation}
\Gamma_{1,0}=\frac{i/l}{\Lambda_{l}(\nu_{1,0})}
\int^\infty_{-\infty}\;d\omega
\frac{g(\omega)\,I_{1,0}(\omega)}{(\omega+Kf_0-\nu_{1,0}-ilD)}.
\label{eq:identity10}
\end{equation}
Since $\Lambda_{l}(\nu_{1,0})=\Lambda_{l}(z_0+2\gamma/l)$ from Eq.
(\ref{eq:numj}), the first factor in
(\ref{eq:identity10}) is singular as $\gamma\rightarrow0^+$. However the
integral is nonsingular in this limit and because of the orthogonality relation
(\ref{eq:Iorthog}) the leading term is zero so the integral is actually ${\cal
O}(\gamma)$. The vanishing of the integral as $\gamma\rightarrow0^+$ cancels
the singularity from $\Lambda_{l}(\nu_{1,0})$ and $\Gamma_{1,0}$ is
well-behaved.

The remaining integrals do exhibit pinching singularities which can be
evaluated with the partial fraction procedure illustrated above. Rewriting
$p_2$ to isolate these singular terms we find for $D=0$:
\begin{eqnarray}
p_2&=&-(2\pi i K  l)\left\{f_l\,<\tilde{\psi},h_{2,1}>
+f_{3l}^\ast\Gamma_{3,0}<\tilde{\psi},h_{2,0}^\ast>+\;\;{\rm nonsingular
\;\;terms}\right\}
\label{eq:p2psing}
\end{eqnarray}
with
\begin{eqnarray}
<\tilde{\psi},h_{2,1}>&=&-2\pi K\,f_{3l}^\ast\Gamma_{3,0}\,
\int_{-\infty}^\infty\,
\frac{d\omega\,\tilde{\psi}(\omega)^\ast\,{\psi}(\omega)^\ast}
{\omega+K\,f_0-(z_0+i\gamma/l)}+\;\;{\rm nonsingular \;\;terms}
\label{eq:t1}\\
&&\nonumber\\
<\tilde{\psi},h_{2,0}^\ast>&=&
-2\pi K\,f_{l}\,
\int_{-\infty}^\infty\,
\frac{d\omega\,\tilde{\psi}(\omega)^\ast\,{\psi}(\omega)^\ast}
{\omega+K\,f_0-z_0^\ast}.\label{eq:t2}
\end{eqnarray}
A partial fraction expansion of each of these integrals allows an explicit
evaluation of the singular behavior
\begin{eqnarray}
\int_{-\infty}^\infty\,
\frac{d\omega\,\tilde{\psi}(\omega)^\ast\,{\psi}(\omega)^\ast}
{\omega+K\,f_0-(z_0+i\gamma/l)}&=&
-\frac{il^2f_l\;[{\rm Im}(f_l)+{\cal O}(\gamma)]}
{3\,|f_l|^2\Lambda'_{l}(z_0)\,\gamma^2}\label{eq:sing1}\\
&&\nonumber\\
\int_{-\infty}^\infty\,
\frac{d\omega\,\tilde{\psi}(\omega)^\ast\,{\psi}(\omega)^\ast}
{\omega+K\,f_0-z_0^\ast}&=&
\frac{il^2f_l\;[{\rm Im}(f_l)+{\cal O}(\gamma)]}
{2\,|f_l|^2\Lambda'_{l}(z_0)\,\gamma^2};\label{eq:sing2}
\end{eqnarray}
each integral diverges like $\gamma^{-2}$.

The resulting evaluation of
$p_2$ yields
\begin{eqnarray}
p_2&=&-\frac{2l(\pi\, l\,K\,f_l)^2\,f_{3l}^\ast\,\Gamma_{3,0}\,
[{\rm Im}(f_l)+{\cal O}(\gamma)]}
{3\,|f_l|^2\,\Lambda'_{l}(z_0)\,\gamma^2}+\;\;{\rm nonsingular \;\;terms};
\label{eq:p2final}
\end{eqnarray}
finally, the $\Gamma_{3,0}$ factor is generically nonzero and can be evaluated
from (\ref{eq:identity1}) as
\begin{equation}
\Gamma_{3,0}=
-2(\pi\,K\,f_l^\ast)^2\,\frac{\Lambda''_{l}(z_0)}{\Lambda_{3l}(z_0)}.
\end{equation}
Hence $p_2\sim\gamma^{-2}$ for the special case $f_{2l}=0$ when $p_1$ is
nonsingular. However, if $f_{2l}$ {\em and} $f_{3l}$ both vanish,
then the divergent terms
in (\ref{eq:p2final}) disappear and both $p_1$ and $p_2$
are nonsingular.

These conclusions are generalized below.  In Proposition
\ref{prop:f2l0} we show that when $f_{2l}=0$ but $f_{3l}\neq0$, then
$p_j\sim\gamma^{-2j+2}$ describes the strongest possible singularity of $p_j$.
In Proposition \ref{prop:kur} we find that  the coefficients $p_j$ are
nonsingular at every order if $f_{nl}=0$ for all $n>1$.

The implications of the $p_2$ divergence for the scaling of the amplitude,
$\gamma^\beta\,r(\gamma t)$,
can be seen by rewriting the radial equation (\ref{eq:radial}) to allow for
both the third and fifth order terms:
\begin{equation}
\frac{d r}{d\tau}=\gamma r+ {\rm Re}(p_1) \gamma^{2\beta-1}r^3
+ {\rm Re}(p_2) \gamma^{4\beta-1}r^3.\label{eq:fifth}
\end{equation}
When $p_1$ is nonsingular and $p_2\sim\gamma^{-2}$, then we must take
$\beta=3/4$ to obtain an expansion that is finite through fifth order terms.
In this case however the general singularity result $p_j\sim\gamma^{-2j+2}$
in Proposition \ref{prop:f2l0} does not exclude higher order terms that
would require a still larger value of $\beta$; this point is discussed
further in Section \ref{sec:newscaling} below.

\section{Singularity structure of the expansion to all orders}\label{sec:sing}
Our goal is a systematic understanding of the singularities in $p_j$
to all orders
in the amplitude expansion. These singularities are smoothed by the noise
(cf. the singularity in (\ref{eq:sing})),
and in the discussion below we always set $D=0$.
In this analysis, it is useful to introduce an index which keeps track of the
worst case singularity for each coefficient.

\subsection{Definition of the index}

The pinching singularities at every order occur in integrands of a standard
form. Let
\begin{equation}
D_n(\alpha,\omega)\equiv\frac{1}{(\omega-\alpha_1)(\omega-\alpha_2)
\cdots(\omega-\alpha_n)}
\label{eq:Ddef}
\end{equation}
for $n>0$ where  $\alpha\equiv(\alpha_1,\ldots,\alpha_n)$ and define
$D_0(\alpha,\omega)\equiv1$. Evaluating $p_j$ for $j\geq1$ involves integrals
of the form $\int d\omega\,\phi(\omega){\cal G}(\omega)$ where
\begin{equation}
{\cal G}(\omega)=D_m(\beta,\omega)^\ast\,D_n(\alpha,\omega)
\label{eq:indatom}
\end{equation}
with $m+n\geq1$. Here $\phi(\omega)$ denotes a smooth function assumed to be
well behaved as $\gamma\rightarrow0$ and to vanish for
$|\omega|\rightarrow\infty$. For example, $\phi(\omega)=g(\omega)$ is often the
smooth function of interest. The poles in (\ref{eq:indatom}) are given by
\begin{eqnarray}
\alpha_j&=&z_0-Kf_0+i\gamma\delta_j\hspace{0.5in}j=
1,\ldots,n\label{eq:poles1}\\
\beta^\ast_j&=&
z_0^\ast-Kf_0-i\gamma\zeta_j\hspace{0.5in}j=1,\ldots,m;\label{eq:poles2}
\end{eqnarray}
hence they lie along the vertical line $\omega={\rm Re }(z_0)-Kf_0$ at
locations determined by the numbers $\delta_j\geq0$ and $\zeta_j\geq0$ which
are assumed to be independent of $\gamma=l\,{\rm Im}(z_0)$.

The {\em index} of ${\cal G}(\omega)$ in (\ref{eq:indatom}) is defined to be
\begin{equation}
{\mbox{\rm Ind }} [{\cal G}]\equiv m+n-1.\label{eq:inddef}
\end{equation}
We extend this definition to assign an index if ${\cal G}(\omega)$ is
multiplied
by a smooth function $\phi(\omega)$ as described above
\begin{equation}
{\mbox{\rm Ind }} [\phi(\omega){\cal G}(\omega)]
\equiv{\mbox{\rm Ind }} [{\cal G}(\omega)].\label{eq:inddefb}
\end{equation}

Since we assume $m+n\geq1$ in (\ref{eq:indatom}), ${\mbox{\rm Ind }}
[\phi\,{\cal G}]\geq0$. If $mn\neq0$, then as ${\gamma\rightarrow0^+}$, the
integral $\int d\omega\,\phi(\omega){\cal G}(\omega)$ may diverge due to a
pinching singularity at $\omega={\rm Re }(z_0)-Kf_0$. The smooth functions
$\phi(\omega)$ are always assumed to be sufficiently well behaved at large
$|\omega|$ that the only possible divergence is due to the pinching
singularity. The index of $\phi\,{\cal G}$ is simply related to the maximum
possible strength of this divergence.

\begin{prop}\label{prop:sing} For ${\cal G}(\omega)$ in \mbox{\rm
(\ref{eq:indatom})} with $m+n\geq1$ and $\phi(\omega)$ a smooth function as
described above,
the integral of $\phi\,{\cal G}$ satisfies
\begin{equation}
\lim_{{\gamma\rightarrow0^+}}\; \left|\gamma^{J}\;
\int^\infty_{-\infty}\,d\omega\,\phi(\omega)\;{\cal
G}(\omega)\right|<\infty\label{eq:sing0}
\end{equation}
with  $J={\mbox{\rm Ind }} [\phi\,{\cal G}]$. If $J$ is replaced by $J-1$,
then the limit \mbox{\rm (\ref{eq:sing0})} diverges in general unless
$mn=0$ in which case the limit is zero for any $J>0$.
\end{prop}
\noindent {\em {\bf Proof}.}
\begin{quote} See Appendix \ref{app:proof}.
{\bf $\Box$}\end{quote}

Our application of Prop. \ref{prop:sing} to the recursion relations for
$I_{m,j}$ requires a further generalization of the index to allow for sums of
functions with well-defined indices and products of $\phi\,{\cal G}$ with
singular functions of $\gamma$. In each case, the generalized index is defined
so that Eq. (\ref{eq:sing0}) remains true, i.e. the index for the composite
function gives the {\em maximal} possible divergence of its integral.

First, let $G_1(\omega)$ and $G_2(\omega)$ denote functions with well-defined
indices, ${\mbox{\rm Ind }} [{ G}_1]\geq{\mbox{\rm Ind }} [{ G}_2]$. The index
of their sum is defined as the largest of the two individual indices:
\begin{equation}
{\mbox{\rm Ind }} [{ G}_1+{G}_2]\equiv{\mbox{\rm Ind }} [{ G}_1].\label{eq:sum}
\end{equation}
Clearly, ${\mbox{\rm Ind }} [{ G}_1]$ gives the maximal possible divergence of
$\int\,d\omega\,(G_1+G_2)$.
Secondly, let $q(\gamma)$ denote a function of $\gamma$ with the asymptotic
behavior
\begin{equation}
\lim_{\gamma\rightarrow0^+}\;q(\gamma)\sim\gamma^{-\nu},
\end{equation}
then the index of $q(\gamma)\,{ G_1}(\omega)$ is defined by
\begin{equation}
{\mbox{\rm Ind }} [q\,{ G_1}]\equiv {\mbox{\rm Ind }}
[{G_1}]+\nu.\label{eq:prod}
\end{equation}
This completes the definition of the index.

By applying (\ref{eq:sum}) and (\ref{eq:prod}) the
indices of $I_{m,j}$ and $h_{m,j}$ are determined from the recursion relations;
some simple
examples are provided in section \ref{sec:examples}. We stress again that
when applied to a composite function $G(\omega)$ the estimate in
(\ref{eq:sing0}) does not necessarily determine the true singularity, but only
an upper bound on
the possible divergence of the integral. In many examples considered below this
upper bound is actually realized, but not always.

There are two immediate implications of the index definition worth noting.
First, complex conjugation doesn't change the index
\begin{equation}
{\mbox{\rm Ind }} [{ G}]={\mbox{\rm Ind }} [{ G}^\ast].\label{eq:cc}
\end{equation}
Secondly, if $G(\omega)$ has a well defined index, then dividing $G$ by
$(\omega-\alpha)$ or $(\omega-\beta^\ast)$
simply increases the index of $G$ by one:
\begin{equation}
{\mbox{\rm Ind }} [{ G}/(\omega-\alpha)]=
{\mbox{\rm Ind }} [{ G}/(\omega-\beta^\ast)]=
{\mbox{\rm Ind }} [{ G}]+1.\label{eq:divide}
\end{equation}
Here $\alpha, \beta$ introduce additional poles as defined in (\ref{eq:poles1})
- (\ref{eq:poles2}) and thereby increase the index of every term in $G$ by one;
this implies (\ref{eq:divide}).

The second observation (\ref{eq:divide}) provides a useful
relation between the divergence of $\Gamma_{m,j}$ and the divergence of
$<\tilde{\psi},h_{m,j}>$. Suppose that $\gamma^{-\nu}$, $\nu>0$, denotes the
actual divergence of $\Gamma_{m,j}$.  Then, from (\ref{eq:identity1}) this must
be due to certain terms in $I_{m,j}$; in addition when $m=1$ then
$\Lambda_{l}(\nu_{1,j})$ is of order
$\gamma$ for small $\gamma$, which increases the divergence by one.
{}From (\ref{eq:identity2}),
$<\tilde{\psi},h_{m,j}>$ is given by
\begin{eqnarray}
<\tilde{\psi},h_{m,j}>&=&\frac{-i}{ml\,\Lambda'_{l}(z_0)}
\int^\infty_{-\infty}\;
\frac{d\omega\,g(\omega)\,I_{m,j}(\omega)}
{(\omega+Kf_0-z_0)(\omega+Kf_0-\nu_{m,j})}\label{eq:int2}\\
&&\nonumber\\
&&\hspace{0.25in}+
\frac{K f_{ml}^\ast\,\Gamma_{m,j}}{\Lambda'_{l}(z_0)}
\int^\infty_{-\infty}\;
\frac{d\omega\,g(\omega)}{(\omega+Kf_0-z_0)(\omega+Kf_0-\nu_{m,j})},
\nonumber
\end{eqnarray}
and we conclude that its divergence cannot be worse than
$\gamma^{-(\nu+1-\delta{m,1})}$.
The second integral
in  (\ref{eq:int2}) has no pinching singularity and the singularity
of the second term is determined by $\Gamma_{m,j}\sim\gamma^{-\nu}$.
The first term in (\ref{eq:int2}) involves an integral whose integrand
differs from the integrand in (\ref{eq:identity2}) only by the extra factor
$(\omega+Kf_0-z_0)$ in the denominator.
{}From (\ref{eq:divide}) this factor simply increases the index of every term
in $I_{m,j}(\omega)/{(\omega+Kf_0-\nu_{m,j})}$. In particular,  the terms that
determined the singularity of $\Gamma_{m,j}\sim\gamma^{-\nu}$ in the first
place have their index increased by 1. Thus we can infer the divergence of the
first term in (\ref{eq:int2}) by adding one to $\nu$ and subtracting off the
effect of $\Lambda_{ml}(\nu_{m,j})$ if $m=1$;
this proves that the singularity of $<\tilde{\psi},h_{m,j}>$ is no
worse than $\gamma^{-(\nu+1-\delta{m,1})}$.  Note that if $\Gamma_{m,j}$ is not
singular ($\nu=0$) then the same reasoning shows that $<\tilde{\psi},h_{m,j}>$
will also be nonsingular.
In summary, we have proven
\begin{prop}\label{prop:prop2}
If the asymptotic behavior of $\Gamma_{m,j}$
satisfies
\begin{equation}
\lim_{\gamma\rightarrow0^+}\gamma^{\nu}|\Gamma_{m,j}|<\infty
\label{eq:prop2a}
\end{equation}
for an integer $\nu>0$, then
\begin{equation}
\lim_{\gamma\rightarrow0^+}\gamma^{\nu'}|<\tilde{\psi},h_{m,j}>|<\infty
\label{eq:pjint}
\end{equation}
with
\begin{equation}
{\nu'}=\left\{\begin{array}{cc}
\nu& m=1\\
\nu+1&m>1.\end{array}\right.\label{eq:prop2c}
\end{equation}
If $\nu=0$ in \mbox{\rm (\ref{eq:prop2a})}, then
\begin{equation}
\lim_{\gamma\rightarrow0^+}|<\tilde{\psi},h_{m,j}>|<\infty.\label{eq:prop2e}
\end{equation}
\end{prop}
This result is useful because $\nu'$ in (\ref{eq:prop2c}) often turns out
to be less than the index of the integrand of $<\tilde{\psi},h_{m,j}>$ and
thus the Proposition can provide sharper estimates of the actual divergences.

It is very important to recognize that the conclusions in (\ref{eq:pjint})
and (\ref{eq:prop2e})
do not in general apply to $<\tilde{\psi},h_{m,j}^\ast>$. Even though
$h_{m,j}$ and $h_{m,j}^\ast$ have the same index, the most singular terms in
$<\tilde{\psi},h_{m,j}>$ are not necessarily the most singular terms in
$<\tilde{\psi},h_{m,j}^\ast>$. For $<\tilde{\psi},h_{m,j}^\ast>$,
if we wish to avoid explicitly evaluating the integral, our
only means of estimating the divergence is by calculating the index of the
integrand.

\subsection{Calculation of the index}\label{sec:examples}

These definitions allow the index of $I_{m,j}$ and $h_{m,j}$ to be calculated
recursively to all orders. We illustrate this index calculation for the first
few levels of Table I; our most general results are stated in Theorem
\ref{thm:main} and
Proposition \ref{prop:f2l0}.

At the top of Table I we have the eigenfunctions defined by linear theory,
these have a single pole in the upper half plane so
\begin{equation}
{\mbox{\rm Ind }} [\psi(\omega)]=
{\mbox{\rm Ind }} [\tilde{\psi}(\omega)]=0.\label{eq:efcni}
\end{equation}
Next we consider $I_{2,0}$ and $h_{2,0}$; from (\ref{eq:i20}) and
(\ref{eq:cmcoeff}) we calculate
\begin{equation}
{\mbox{\rm Ind }} [I_{2,0}(\omega)]=0\hspace{1.0in}
{\mbox{\rm Ind }} [h_{2,0}(\omega)]=1.\label{eq:linind}
\end{equation}
The index of $I_{2,0}$ follows from (\ref{eq:efcni}), while the expression for
$h_{2,0}$ has two terms; the first has index 1 and the second has index 0,
hence the index of $h_{2,0}$ is 1.

Since $h_{2,0}$ only has poles in the upper half
plane, the integrals $\Gamma_{2,0}$ and $<\tilde{\psi},h_{2,0}>$ have no
pinching singularities and remain finite as $\gamma\rightarrow0^+$:
\begin{eqnarray}
\Gamma_{2,0}&\sim&\gamma^{0}\hspace{1.0in}
<\tilde{\psi},h_{2,0}>\sim\gamma^{0}.\label{eq:exsing}
\end{eqnarray}
Note that the integrand in (\ref{eq:exsing}) has index equal to 2, but there is
no
divergence, this illustrates the second part of Proposition \ref{prop:prop2}.
By contrast the integrand of $<\tilde{\psi},\psi^\ast>$ has a smaller index
than
$<\tilde{\psi},h_{2,0}>$ but yields a singular integral (\ref{eq:sing}) as
$\gamma\rightarrow0^+$:
\begin{equation}
<\tilde{\psi},\psi^\ast>\sim\gamma^{-1}.\label{eq:cubdiv}
\end{equation}

The fifth order coefficient $p_2$ provides an instructive exercise in applying
the index to estimate the singularity of a more complicated expression. The
general form of the coefficient (not assuming $f_{2l}=0$) is
\begin{eqnarray}
p_2&=&-(2\pi i K  l)\left\{f_l\left[<\tilde{\psi},h_{2,1}>+
\Gamma_{1,0}^\ast<\tilde{\psi},h_{2,0}>\right] +
f_{2l}\,\Gamma_{2,0}^\ast<\tilde{\psi},h_{3,0}> \right.\\
&&\hspace{0.75in}\left.+f_{2l}^\ast
\left[\Gamma_{2,1}<\tilde{\psi},\psi^\ast>+
\Gamma_{2,0}<\tilde{\psi},h_{1,0}^\ast>\right]+
f_{3l}^\ast\Gamma_{3,0}<\tilde{\psi},h_{2,0}^\ast>\right\}\nonumber
\end{eqnarray}
where
\begin{eqnarray}
I_{3,0}&=&2\pi i K  (3l)
\;[f_l^\ast\; h_{2,0}+f_{2l}^\ast\,\psi\,\Gamma_{2,0}]\label{eq:I30b}\\
I_{1,0}&=&(2\pi i K  l)\;{\rm P}_\perp
\left\{f_{l}h_{2,0}+f_{2l}^\ast\,\psi^\ast\,\Gamma_{2,0}\right\}
\label{eq:I10b}\\
I_{2,1}&=&2p_1h_{2,0}+2\pi i K  (2l)\,
\left[f_lh_{3,0}+f_l^\ast(h_{1,0}+\Gamma_{1,0}\psi)+
f_{3l}^\ast\Gamma_{3,0}\psi^\ast\right].
\end{eqnarray}

We can immediately estimate the integral $<\tilde{\psi},h_{2,0}^\ast>$ from
the fact that the integrand has index 2
\begin{equation}
<\tilde{\psi},h_{2,0}^\ast>\sim\gamma^{-2}.
\end{equation}

The two terms in $I_{3,0}$ have indices of 1 and 0, respectively, and the poles
of $I_{3,0}$ are all in the upper half plane. Writing out the expression in
(\ref{eq:I10b})
\begin{equation}
{\rm P}_\perp
\left\{f_{l}h_{2,0}+f_{2l}^\ast\,\psi^\ast\,\Gamma_{2,0}\right\}=
f_{l}h_{2,0}+f_{2l}^\ast\,\psi^\ast\,\Gamma_{2,0}-
\psi\,<\tilde{\psi},[f_{l}h_{2,0}+f_{2l}^\ast\,\psi^\ast\,\Gamma_{2,0}]>,
\end{equation}
we find terms of index 1, 0, and 1, respectively. The third term is already
singular due to $<\tilde{\psi},\psi^\ast>$ and the second term has a pole
in the lower half plane. These observations imply
\begin{eqnarray}
{\mbox{\rm Ind }} [I_{3,0}(\omega)]&=&1\hspace{1.0in}
{\mbox{\rm Ind }} [I_{1,0}(\omega)]\leq1,\label{eq:g10}
\end{eqnarray}
and from (\ref{eq:identity1})
\begin{equation}
\Gamma_{1,0}\sim\gamma^{-2}\hspace{1.0in}
\Gamma_{3,0}\sim\gamma^{0}.
\end{equation}
The inequality in (\ref{eq:g10}) reflects the fact that there are two
terms of index 1 and we have not ruled out the possibility of a  cancellation.
The indices of $h_{3,0}$ and $h_{1,0}$ are now easily evaluated from
(\ref{eq:identity2})
\begin{eqnarray}
{\mbox{\rm Ind }} [h_{3,0}(\omega)]&=&2\hspace{1.0in}
{\mbox{\rm Ind }} [h_{1,0}(\omega)]\leq2,\label{eq:h10}
\end{eqnarray}
and the corresponding integrals are estimated following (\ref{eq:pjint})
\begin{eqnarray}
<\tilde{\psi},h_{3,0}>&\sim&\gamma^{0}\hspace{1.0in}
<\tilde{\psi},h_{1,0}>\sim\gamma^{-2}.\label{eq:h10int}
\end{eqnarray}
For $p_2$ we also need to estimate $<\tilde{\psi},h_{1,0}^\ast>$ and this
is done by noting from (\ref{eq:h10}) that the integrand has an index of (at
most) 3 so
\begin{equation}
<\tilde{\psi},h_{1,0}^\ast>\sim\gamma^{-3}.
\end{equation}

The five terms in $I_{2,1}$ all have index 2 except $\Gamma_{3,0}\psi^\ast$
which has index 0, hence
\begin{equation}
{\mbox{\rm Ind }} [I_{2,1}(\omega)]\leq2.
\end{equation}
The terms involving $h_{1,0}$ and $\psi^\ast$ have poles in the lower half
plane and therefore produce pinching singularities in $\Gamma_{2,1}$. By
comparing
(\ref{eq:identity1}) for $\Gamma_{2,1}$ to (\ref{eq:cubdiv}) and
(\ref{eq:h10int}) we can estimate a maximum singularity of
\begin{equation}
\Gamma_{2,1}\sim\gamma^{-2}
\end{equation}
which in turn implies ${\mbox{\rm Ind }} [h_{2,1}(\omega)]\leq3,$ and
\begin{equation}
<\tilde{\psi},h_{2,1}>\sim\gamma^{-3}\label{eq:term1}
\end{equation}
from (\ref{eq:identity2}) and (\ref{eq:pjint}), respectively.

The result in (\ref{eq:term1}) gives the divergence of the first term in $p_2$,
the asymptotic behavior of the remaining five terms can be evaluated
from the foregoing estimates:
\begin{eqnarray}
\Gamma_{1,0}^\ast<\tilde{\psi},h_{2,0}>&\sim&\gamma^{-2}\\
\Gamma_{2,0}^\ast<\tilde{\psi},h_{3,0}>&\sim&\gamma^{0}\\
\Gamma_{2,1}<\tilde{\psi},\psi^\ast>&\sim&\gamma^{-3}\\
\Gamma_{2,0}<\tilde{\psi},h_{1,0}^\ast>&\sim&\gamma^{-3}\\
\Gamma_{3,0}<\tilde{\psi},h_{2,0}^\ast>&\sim&\gamma^{-2}.
\end{eqnarray}
Hence barring some accidental cancellation among the most divergent terms we
expect $p_2\sim\gamma^{-3}$ for the general case when $f_{2l}\neq0$; more
precisely we have proved
\begin{equation}
\lim_{\gamma\rightarrow0^+}\gamma^3|p_2|<\infty.\label{eq:p2gen}
\end{equation}

\subsection{The main result}

Our main result on the singularity structure of the amplitude expansions can
now be proved. In the preceeding sections, we have calculated the
indices of $I_{m,j}$ and $h_{m,j}$ through the first two levels of Table I, and
determined the singular behavior of the cubic and fifth order coefficients.
This information is now inferred for the entire theory to all orders.
\begin{theorem}\label{thm:main}
For $j\geq1$, the singularities of the coefficients in the amplitude expansion
satisfy
\begin{equation}
\lim_{{\gamma\rightarrow0^+}}\;
\gamma^{2j-1}\;\left|p_j\right|<\infty.\label{eq:pjasy}
\end{equation}
For $m\geq1$ and $j\geq0$, the indices of $I_{m,j}$ and $h_{m,j}$ obey
\begin{eqnarray}
{\mbox{\rm Ind }} \left[I_{m,j}\right]&\leq&
J_{m,j}+2\delta_{m,1}.\label{eq:keyind}\\
{\mbox{\rm Ind }} \left[h_{m,j}\right]&\leq&
J_{m,j}+1+2\delta_{m,1}\label{eq:hind}
\end{eqnarray}
where $J_{m,j}\equiv m+2j-2$, and the integrals in {\rm (\ref{eq:pj})} satisfy
\begin{eqnarray}
\lim_{{\gamma\rightarrow0^+}}\;\gamma^{J_{m,j}+3\delta_{m,1}}\;
\left|\Gamma_{m,j}
\right|&<&\infty\label{eq:gmklasy}\\
\lim_{{\gamma\rightarrow0^+}}\;\gamma^{J_{m,j}+1+2\delta_{m,1}}\;
\left|<\tilde{\psi},h_{m,j}>\right|
&<&\infty.\label{eq:intest}
\end{eqnarray}
\end{theorem}

We are only able to prove the results in (\ref{eq:keyind}) - (\ref{eq:hind}) as
upper bounds on the index, but we expect them to hold as equalities in most
cases.
Note that the bounds in (\ref{eq:gmklasy}) - (\ref{eq:intest}) give
weaker divergences than one would estimate simply from the index
of the integrand as calculated using (\ref{eq:keyind}) - (\ref{eq:hind}).
For example, from (\ref{eq:hind}) the integrand of $<\tilde{\psi},h_{m,j}>$ has
index $J_{m,j}+2+2\delta_{m,1}$, but the bound in (\ref{eq:intest}) assures us
that the divergence of this integral will always be at least one power
of $\gamma$ less. There is no contradiction since the index only gives a strict
upper limit on divergence exhibited by a given  integral.

\noindent  {\em {\bf Proof.}}\begin{quote}
\begin{enumerate}

\item The relations (\ref{eq:pjasy}) - (\ref{eq:intest}) have been checked
explicitly for  $I_{2,0}$, $h_{2,0}$, $\Gamma_{2,0}$, and $p_1$ in the analysis
of section \ref{sec:p1}. In addition, in section \ref{sec:examples}
 they have been verified for
$I_{1,0}$, $h_{1,0}$, $\Gamma_{1,0}$, $I_{3,0}$, $h_{3,0}$, $\Gamma_{3,0}$,
$I_{2,1}$, $h_{2,1}$, $\Gamma_{2,1}$, and $p_2$. This proves the theorem
for the first two levels of Table I.

\item  We extend (\ref{eq:pjasy}) - (\ref{eq:intest}) to all $p_j$,
and $\{h_{m,j}$, $I_{m,j}\}$ by induction using the recursion relations.
Assume that (\ref{eq:pjasy}) - (\ref{eq:intest}) are true
down to some arbitrary level of Table I, and consider what the recursion
relations imply for the coefficients evaluated at the next level.
Let $h_{m',j'}$ and $I_{m',j'}$ denote coefficients at the next level that
can be evaluated from lower order quantities which satisfy (\ref{eq:pjasy}) -
(\ref{eq:intest}). We first
consider the various recursion relations for $h_{m',j'}$ and $I_{m',j'}$, and
prove  that if (\ref{eq:pjasy}) - (\ref{eq:intest}) hold on
the right hand side of these relations, then (\ref{eq:keyind}) -
(\ref{eq:intest}) will also hold for the coefficients obtained on the left.
We organize this part of the proof by noting the
importance of establishing two relations for $I_{m',j'}$:
\begin{enumerate}
\item  the index identity (\ref{eq:keyind})
\begin{equation}
{\mbox{\rm Ind }} \left[I_{m',j'}\right]\leq
J_{m',j'}+2\delta_{m',1}.\label{eq:keyind2}
\end{equation}
\item the estimate
\begin{equation}
\lim_{{\gamma\rightarrow0^+}}\;\gamma^{J_{m',j'}+2\delta_{m',1}}
\left|\int^\infty_{-\infty}\;d\omega
\frac{g(\omega)\,I_{m',j'}(\omega)}{(\omega-\alpha)}\right|<\infty
\label{eq:indstep}
\end{equation}
where $\alpha$ is any pole of the form described in (\ref{eq:poles1})
(in the upper half plane). Note that (\ref{eq:keyind2}) implies that the
integrand in (\ref{eq:indstep}) has
an index of $J_{m',j'}+2\delta_{m',1}+1$; thus (\ref{eq:indstep}) gives a
divergence which is less by one power of $(1/\gamma)$ than would be ``naively''
predicted using
the index of $I_{m',j'}$ in (\ref{eq:keyind2}).
\end{enumerate}
If we can prove (\ref{eq:keyind2}) and (\ref{eq:indstep}) for $I_{m',j'}$,
then the remaining
properties (\ref{eq:hind}) - (\ref{eq:intest}) are easily obtained as
follows. First,
(\ref{eq:gmklasy}) follows from (\ref{eq:indstep})
and the identity (\ref{eq:identity1}). Next (\ref{eq:hind}) follows
by applying (\ref{eq:keyind2}) and  (\ref{eq:gmklasy}) to the identity
(\ref{eq:identity2}).
Finally (\ref{eq:intest}) follows from  Proposition \ref{prop:prop2} and
(\ref{eq:gmklasy}).
Thus the crux of the matter is to
verify (\ref{eq:keyind2}) and (\ref{eq:indstep}); this is done using
the recursion relations for  $I_{m',j'}$, i.e. using (\ref{eq:i1j}) -
(\ref{eq:imj}).

\item {\em Verification of {\rm (\ref{eq:keyind2})}}.

The index of each term on the right hand side of the
recursion relations for $I_{m',j'}(\omega)$ can be evaluated by applying
(\ref{eq:pjasy}) - (\ref{eq:intest}); this exercise shows that all of these
terms have an index less than or equal to $J_{m',j'}+2\delta_{m',1}$.
This establishes (\ref{eq:keyind2}) for $I_{m',j'}$. A few examples from
(\ref{eq:i1j}) illustrate these index calculations.
\begin{enumerate}

\item  The terms in $I_{1,j}$ that depend on $p_j$ have the form
$[(2+k)p_{j-k}+(1+k)p_{j-k}^\ast]\,h_{1,k}$; from (\ref{eq:hind}) we have
${\mbox{\rm Ind }} \left[h_{1,k}\right]\leq2k+2$ and from (\ref{eq:pjasy}) the
singularity of $p_{j-k}$ is determined, hence
\begin{equation}
{\mbox{\rm Ind }} \left[p_{j-k}\,h_{1,k}\right]\leq2(j-k)-1+(2k+2)=2j+1
\label{eq:index1}
\end{equation}
which is consistent with (\ref{eq:keyind2}) for $(m',j')=(1,j)$.

\item Next consider the terms
\begin{equation}
\sum_{m=3}^{j+2}f_{ml}^\ast\,\sum_{n=0}^{j-m+2}\,{\rm P}_\perp
[h_{m-1,n}^\ast\Gamma_{m,j-m-n+2}]
\label{eq:secondex}
\end{equation}
which expand to
\begin{equation}
{\rm P}_\perp
[h_{m-1,n}^\ast\Gamma_{m,j-m-n+2}]=
\Gamma_{m,j-m-n+2}\,[h_{m-1,n}^\ast-
<\tilde{\psi},h_{m-1,n}^\ast>\,\psi].\label{eq:index3}
\end{equation}
The index of each term can be evaluated
\begin{eqnarray}
{\mbox{\rm Ind }} \left[h_{m-1,n}^\ast\Gamma_{m,j-m-n+2}\right]&\leq&2j
\label{eq:index3a}\\
{\mbox{\rm Ind }}
\left[<\tilde{\psi},h_{m-1,n}^\ast>\,\Gamma_{m,j-m-n+2}\,\psi\right]
\label{eq:index3b}
&\leq&2j+1;
\end{eqnarray}
thus ${\mbox{\rm Ind }} \left[{\rm P}_\perp
[h_{m-1,n}^\ast\Gamma_{m,j-m-n+2}]\right]\leq2j+1$
which is consistent with (\ref{eq:keyind2}) for $(m',j')=(1,j)$.
In this example, both indicies are obtained from  (\ref{eq:hind}) and
(\ref{eq:gmklasy}). The divergence of
$\Gamma_{m,j-m-n+2}\sim\gamma^{-(2j-m-2n+2)}$ is estimated from
(\ref{eq:gmklasy}),
and (\ref{eq:hind}) gives ${\mbox{\rm Ind }} \left[h_{m-1,n}^\ast\right]\leq
m+2n-2$. Hence the integrand of $<\tilde{\psi},h_{m-1,n}^\ast>$ has an index no
greater than $m+2n-1$, and we
conclude that the worst possible divergence is
$<\tilde{\psi},h_{m-1,n}^\ast>\sim\gamma^{-(m+2n-1)}$.
These estimates immediately
imply (\ref{eq:index3a}) - (\ref{eq:index3b}).

\item A more subtle example is the term
${\rm P}_\perp\,h_{2,j}$ in (\ref{eq:i1j}); from (\ref{eq:projop}) we have that
the index
of ${\rm P}_\perp\,h_{2,j}=h_{2,j}-\psi<\tilde{\psi},h_{2,j}>$
 is equal to the largest index
obtained from $h_{2,j}$ and
$<\tilde{\psi},h_{2,j}>\,\psi(\omega)$. These terms have indices
\begin{equation}
{\mbox{\rm Ind }} \left[h_{2,j}\right]\leq2j+1\hspace{0.5in}
{\mbox{\rm Ind }} \left[<\tilde{\psi},h_{2,j}>\,\psi(\omega)\right]\leq2j+1,
\label{eq:index2}
\end{equation}
so ${\mbox{\rm Ind }} \left[{\rm P}_\perp\,h_{2,j}\right]\leq2j+1$ which is
consistent with (\ref{eq:keyind2}) for $(m',j')=(1,j)$. The first index in
(\ref{eq:index2}) comes from (\ref{eq:hind}) and the second follows from
(\ref{eq:intest}) and (\ref{eq:linind}). This example is delicate because
(\ref{eq:index2}) requires that we use (\ref{eq:intest}) to estimate
$<\tilde{\psi},h_{2,j}>\sim\gamma^{-(2j+1)}$
rather than (\ref{eq:hind}) as in the previous calculation (\ref{eq:index3b}).
\end{enumerate}
Proceeding in this way the index of every term in the recursion relations
(\ref{eq:i1j}) - (\ref{eq:imj})
can be calculated and (\ref{eq:keyind2}) verified.

\item {\em Verification of {\rm (\ref{eq:indstep})}}.

We prove (\ref{eq:indstep}) by integrating the recursion relations for
$I_{m',j'}(\omega)$ and examining the resulting recursion relations for
\begin{equation}
\int^\infty_{-\infty}\;d\omega
\frac{g(\omega)\,I_{m',j'}(\omega)}{(\omega-\alpha)}.
\label{eq:crux}
\end{equation}
The right hand side of the recursion relations (\ref{eq:i1j}) - (\ref{eq:imj})
involve products of $\psi$, $\psi^\ast$, $h_{m,j}$, or $h_{m,j}^\ast$ with
integrals whose singularities can be estimated by applying (\ref{eq:pjasy}) -
(\ref{eq:intest}) as illustrated above. Now when we divide by $(\omega-\alpha)$
and integrate as in (\ref{eq:crux}) we get additional integrals of the
following types
\begin{eqnarray}
\int^\infty_{-\infty}\;d\omega
\frac{g(\omega)\,\psi(\omega)}{(\omega-\alpha)}&\hspace{0.3in}&
\int^\infty_{-\infty}\;d\omega
\frac{g(\omega)\,\psi^\ast(\omega)}{(\omega-\alpha)}\label{eq:inta}\\
&&\nonumber\\
\int^\infty_{-\infty}\;d\omega
\frac{g(\omega)\,h_{m,j}^\ast(\omega)}{(\omega-\alpha)}
&\hspace{0.3in}&\int^\infty_{-\infty}\;d\omega
\frac{g(\omega)\,h_{m,j}(\omega)}{(\omega-\alpha)}.\label{eq:intb}
\end{eqnarray}
The first integral in (\ref{eq:inta}) has no pinching singularity and the
second has index 1 and diverges as $\gamma^{-1}$; the singularity of the
integrals in (\ref{eq:intb}) can be estimated from (\ref{eq:hind}) and
(\ref{eq:intest}), respectively:
\begin{eqnarray}
\lim_{{\gamma\rightarrow0^+}}\;\gamma^{J_{m,j}+2+2\delta_{m,1}}
\left|\int^\infty_{-\infty}\;d\omega
\frac{g(\omega)\,h_{m,j}^\ast(\omega)}{(\omega-\alpha)}\right|&<&\infty
\label{eq:intd}\\
\lim_{{\gamma\rightarrow0^+}}\;\gamma^{J_{m,j}+1 +2\delta_{m,1}}
\left|\int^\infty_{-\infty}\;d\omega
\frac{g(\omega)\,h_{m,j}(\omega)}{(\omega-\alpha)}\right|&<&\infty.
\label{eq:intc}
\end{eqnarray}
With these estimates the terms contributing to (\ref{eq:crux}) can be shown to
diverge at worst like
\begin{equation}
\left(\frac{1}{\gamma}\right)^{J_{m',j'}+2\delta_{m',1}}.
\end{equation}
This establishes (\ref{eq:indstep}) for $I_{m',j'}$, and completes the
induction
argument that (\ref{eq:keyind}) - (\ref{eq:intest}) are propagated by the
recursion relations.

\item Now we finish the proof of the theorem by verifying that the estimates
for the integrals in (\ref{eq:gmklasy}) and (\ref{eq:intest}) imply the
estimate of $p_j$ in (\ref{eq:pjasy}); this is a straightforward application of
(\ref{eq:gmklasy})
and (\ref{eq:intest}) to the right hand side of the recursion relations
(\ref{eq:pj}). A few examples from (\ref{eq:pj})
suffice to illustrate how this is done.
\begin{enumerate}

\item Consider the term $<\tilde{\psi},h_{2,j-1}>$;
from (\ref{eq:intest}) the singularity of this integral is at most
\begin{equation}
\left|<\tilde{\psi},h_{2,j-1}>\right|\sim
\left(\frac{1}{\gamma}\right)^{J_{2,j-1}+1}=
\left(\frac{1}{\gamma}\right)^{2j-1}
\label{eq:ex1}
\end{equation}
which is consistent with (\ref{eq:pjasy}). This example is another case where
the divergence of the integral is best characterized with (\ref{eq:intest})
rather than inferred from the index of the integrand which is $2j$.

\item Other terms involve products of integrals, for example,
$\Gamma_{2,j-1}<\tilde{\psi},\psi^\ast>$. Using (\ref{eq:gmklasy}) and
(\ref{eq:cubdiv}) we determine that the singularity of this product is at most
\begin{equation}
\left|\Gamma_{2,j-1}<\tilde{\psi},\psi^\ast>\right|\sim
\left(\frac{1}{\gamma}\right)^{J_{2,j-1}+1}
\label{eq:ex2}
\end{equation}
which is the same as in (\ref{eq:ex1}).

\item The terms
\begin{equation}
\sum_{m=3}^{j+1}f_{ml}^\ast\sum_{n=0}^{j-m+1}
\Gamma_{m,n}<\tilde{\psi},h_{m-1,j-n-m+1}^\ast>;
\label{eq:exfinal}
\end{equation}
are a final example. Applying (\ref{eq:hind}) and
(\ref{eq:gmklasy}) shows a maximum singularity of
\begin{equation}
\left|\Gamma_{m,n}<\tilde{\psi},h_{m-1,j-n-m+1}^\ast>\right|\sim
\left(\frac{1}{\gamma}\right)^{J_{m,n}+J_{m-1,j-n-m+1}+2}=
\left(\frac{1}{\gamma}\right)^{2j-1}
\label{eq:ex3}
\end{equation}
which is also consistent with (\ref{eq:pjasy}).
\end{enumerate}
In this manner every term in the recursion relation for $p_j$ is shown to
satisfy the general estimate in (\ref{eq:pjasy}).
\end{enumerate}

{\bf $\Box$}\end{quote}

\subsection{Implications of Theorem \ref{thm:main}}

Our discussion in section \ref{sec:p1} of the significance of the coefficient
singularities did not consider terms in the amplitude equation beyond third
order. If all such terms are included then the amplitude equation
in polar variables (\ref{eq:polar}) becomes
\begin{eqnarray}
\dot{A}&=&\gamma A+\sum_{j=1}^\infty{\rm Re}(p_j)A^{2j+1}\\
\dot{\psi}&=&\Omega -\sum_{j=1}^\infty{\rm Im}(p_j)A^{2j}.
\end{eqnarray}
With the asymptotic behavior of the coefficients is given by Theorem
\ref{thm:main}
\begin{equation}
p_j=\frac{[a_j+{\cal O}(\gamma)]}{\gamma^{2j-1}},
\end{equation}
we re-express the dynamics in terms of  the rescaled amplitude
(\ref{eq:newvar}) ${A}(t)=\gamma^\beta r(\gamma^\delta t)$
\begin{eqnarray}
\frac{d r}{d\tau}&=&\gamma^{1-\delta}
r+\sum_{j=1}^\infty\gamma^{2j(\beta-1)-\delta+1}{\rm Re}(a_j)[1+{\cal
O}(\gamma)] r^{2j+1}\label{eq:rdot}\\
\frac{d \psi}{d t}&=&\Omega -\sum_{j=1}^\infty\gamma^{2j(\beta-1)+1}{\rm
Im}(a_j)[1+{\cal O}(\gamma)]r^{2j}.
\end{eqnarray}
Now we see that the singularities of $p_j$ are such that the choices $\delta=1$
and $\beta=1$ remove them to {\em all} orders, leaving an asymptotically
finite system
\begin{eqnarray}
\frac{d r}{d\tau}&=& r+\sum_{j=1}^\infty{\rm Re}(a_j)[1+{\cal O}(\gamma)]
r^{2j+1}\label{eq:reqn}\\
\frac{d \psi}{d t}&=&\Omega -\gamma\sum_{j=1}^\infty{\rm Im}(a_j)[1+{\cal
O}(\gamma)]r^{2j}.\label{eq:psieqn}
\end{eqnarray}

In (\ref{eq:reqn}) there is a further remarkable consequence of the
singularities, namely that the terms at higher order in $r$ no longer
appear to be negligible even for small $\gamma$. This conclusion
is provisional in the sense that we have not proven $a_j\neq0$ at high
order. Nevertheless, our analysis of $p_1$ and $p_2$ shows that in general
$a_1$ and $a_2$ are non-zero, and inspection of the recursion relations
shows no
evidence that the series will truncate. This feature does not affect
the phase evolution (\ref{eq:psieqn}) quite so dramatically. If we assume
a time-independent solution $r=R_{rw}$ of (\ref{eq:reqn}) then the nonlinear
frequency of this state
\begin{equation}
\Omega_{rw}=\Omega -\gamma\sum_{j=1}^\infty{\rm Im}(a_j)[1+{\cal
O}(\gamma)]R_{rw}^{2j}\label{eq:rwave}
\end{equation}
is very close to the linear frequency $\Omega$ for small $\gamma$.
Such solutions are called ``rotating wave'' states as they describe
a solution (\ref{eq:umfdb})
\begin{eqnarray}
\eta^u(\theta,\omega,t)&=&
\left[\rule{0.0in}{0.25in}\gamma^\beta\,R_{rw}e^{-i\Omega_{rw}t}
\Psi(\theta,\omega) +
\gamma^{3\beta}\,R_{rw}^3e^{-i\Omega_{rw}t}\,
h_1(\omega,\gamma^{2\beta}\,R_{rw}^2)\,e^{il\theta}\right.\label{eq:rwsoln}\\
&&\hspace{1.0in}\left.+
\sum_{m=2}^\infty\gamma^{m\beta}\,R_{rw}^me^{-im\Omega_{rw}t}\,
h_m(\omega,\gamma^{2\beta}\,R_{rw}^2)\,e^{iml\theta}\right]
+ {\rm c.c.}\nonumber
\end{eqnarray}
that appears time independent in a rotating frame
$\theta'=\theta-\Omega_{rw}t/l$.

The asymptotic behavior of $\Gamma_{m,j}$ is given by Theorem \ref{thm:main}
as
\begin{equation}
\Gamma_{m,j}=\frac{[b_{m,j}+{\cal O}(\gamma)]}{\gamma^{J_{m,j}+3\delta_{m,1}}}
\end{equation}
where the constants $b_{m,j}$ are independent of $\gamma$. The expansion
of $\Gamma_m$ thus has the asymptotic form
\begin{equation}
\Gamma_{m}(\sigma)=\sum_{j=0}^\infty
\Gamma_{m,j}\,\sigma^j=\left(\frac{1}{\gamma}\right)^{m-2+3\delta_{m,1}}
\sum_{j=0}^\infty[b_{m,j}+{\cal O}(\gamma)]r^{2j}.
\label{eq:asym}
\end{equation}
This motivates the definition of the nonsingular coefficient $\hat{\Gamma}_{m}$
\begin{equation}
\Gamma_{m}(\sigma)\equiv\left(\frac{1}{\gamma}\right)^{m-2+3\delta_{m,1}}
\hat{\Gamma}_{m}(r^{2});
\label{eq:ghat}
\end{equation}
$\hat{\Gamma}_{m}$ will play a role shortly in our discussion of Daido's
order parameter.

\section{Special Cases: $f_{2l}=0$}\label{sec:special}
The general analysis shows that if $f_{2l}\neq0$, then the resulting
singularities
require a rescaling of the amplitude by a factor of $\gamma$. In this section,
we consider some of the possiblities that can occur for couplings with
$f_{2l}=0$. First, we note that the general recursion relations simplify
somewhat when $f_{2l}=0$; the terms proportional to $f_{2l}$ are explicitly
shown in (\ref{eq:pj}) and (\ref{eq:i1j}) - (\ref{eq:imj}), and we simply drop
them.

\subsection{Absence of the cubic singularity: $f_{2l}=0$ and $f_{3l}\neq0$}

When $f_{2l}=0$ so that the cubic coefficient is non-singular, then
if $f_{3l}\neq0$ the first singularity appears in the fifth order
coefficient $p_2\sim\gamma^{-2}$. The general structure of the amplitude
expansion for this situation is characterized by
the following proposition.
\begin{prop}\label{prop:f2l0}
 Suppose that $f_{2l}=0$ and $f_{3l}\neq0$, then for $j\geq1$, the
singularities of the coefficients in the amplitude expansion satisfy
\begin{equation}
\lim_{{\gamma\rightarrow0^+}}\;
\gamma^{2j-2}\;\left|p_j\right|<\infty.\label{eq:pjasysc}
\end{equation}
For $m\geq1$ and $j\geq0$, the indices of $I_{m,j}$ and $h_{m,j}$ obey
\begin{eqnarray}
{\mbox{\rm Ind }} \left[I_{m,j}\right]&\leq&
J_{m,j}+2\delta_{m,1}\label{eq:keyindsc}\\
{\mbox{\rm Ind }} \left[h_{m,j}\right]&\leq&
J_{m,j}+1+2\delta_{m,1}\label{eq:hindsc}
\end{eqnarray}
where $J_{m,j}\equiv m+2j-2$, and the integrals in {\rm (\ref{eq:pj})}
satisfy
\begin{eqnarray}
\lim_{{\gamma\rightarrow0^+}}\;
\gamma^{J_{m,j}+3\delta_{m,1}-(1-\delta_{m,2}\delta_{j,0})}\;
\left|\Gamma_{m,j}\right|&<&\infty\label{eq:gmklasysc}\\
\lim_{{\gamma\rightarrow0^+}}\;\gamma^{J_{m,j}+2\delta_{m,1}}\;
\left|<\tilde{\psi},h_{m,j}>\right|
&<&\infty.\label{eq:intestsc}
\end{eqnarray}
\end{prop}
Compared to the conclusions of Theorem \ref{thm:main} the bounds on the indices
are the same so there is nothing to prove. The estimates of the singularities
in (\ref{eq:pjasysc}) and (\ref{eq:gmklasysc}) - (\ref{eq:intestsc}) are
softened by one factor of $\gamma$. Since $J_{2,0}=0$, there is no singularity
in $\Gamma_{2,0}$ even in the general case and the odd correction term
$\delta_{m,2}\delta_{j,0}$
in (\ref{eq:gmklasysc}) adjusts the exponent to allow for this.  In fact,
$\Gamma_{m,0}$, $m>1$  is nonsingular in general, although we have not used
this.

\noindent
{\em {\bf Proof.}}
\begin{quote}
\begin{enumerate}
\item The results in (\ref{eq:pjasysc}) - (\ref{eq:intestsc}) have been
established by the calculations in sections \ref{sec:p1} and \ref{sec:p2};
they are extended to all orders by  induction. The argument is entirely
analogous to the proof given for
Theorem \ref{thm:main} and we indicate only the necessary modifications and
omit the details.

\item The only adjustment required is that the second crucial identity
(\ref{eq:indstep}) in the
general proof is now replaced by the estimate
\begin{equation}
\lim_{{\gamma\rightarrow0^+}}\;\gamma^{J_{m',j'}+2\delta_{m',1}
-(1-\delta_{m',2}\delta_{j',0})}
\left|\int^\infty_{-\infty}\;d\omega
\frac{g(\omega)\,I_{m',j'}(\omega)}{(\omega-\alpha)}\right|<\infty
\label{eq:indstepsc}
\end{equation}
where $\alpha$ is any pole of the form described in (\ref{eq:poles1})
(in the upper half plane).
The first index identity (\ref{eq:keyind2}) is still appropriate:
\begin{equation}
{\mbox{\rm Ind }} \left[I_{m',j'}\right]\leq
J_{m',j'}+2\delta_{m',1}.\label{eq:keyind2sc}
\end{equation}
With this adjustment the repetition of the previous argument is
straightforward.
\end{enumerate}

{\bf $\Box$}\end{quote}

\subsection{Implications for scaling}\label{sec:newscaling}

The consequences of this Theorem for the scaling of the amplitude are
ambiguous. If we rewrite the radial equation from (\ref{eq:rdot}) with
$\delta=1$
\begin{equation}
\frac{d r}{d\tau}= r+\sum_{j=1}^\infty\gamma^{2j\beta-1}{\rm Re}(p_j)
r^{2j+1},\label{eq:rnewa}
\end{equation}
and then insert the estimated singularity $p_j\sim a_j\gamma^{-2j+2}$ we find
\begin{equation}
\frac{d r}{d\tau}= r+\sum_{j=1}^\infty\gamma^{2j(\beta-1)+1}{\rm Re}(a_j)
r^{2j+1}.\label{eq:rnewb}
\end{equation}
Now the requirement that all the nonlinear terms should be nonsingular,
$2j(\beta-1)+1\geq0$, results in a $j$-dependent estimate on the exponent:
$\beta\geq1-1/2j$. We know the fifth order singularity occurs so from $j=2$ we
have $\beta=3/4$ as discussed previously in Section \ref{sec:p2}. However our
analysis of the recursion relations is not sharp
enough to determine if the leading singularities are actually present at every
order. Consequently, we have a range of possible values of $\beta$
\begin{equation}
\frac{3}{4}\leq\beta\leq 1.\label{eq:range}
\end{equation}

This range can be reduced to $\beta=1$ if we require that the rescaling
also yield
a finite expansion for $\Gamma$ with coefficient singularities as estimated
in (\ref{eq:gmklasysc}). However it must be borne in mind that these
singularities are upper bounds. Our firmest prediction remains of the range
of values in (\ref{eq:range}) since the bound $\beta\geq3/4$ is conclusively
required by our fifth order calculations and the the bound $\beta\leq1$ follows
from our analysis of the general case.

\subsection{Single component couplings}

An additional case of interest is the circumstance which holds for the Kuramoto
model $f(\phi)=\sin\phi$; a coupling that lacks all harmonics of the Fourier
component $f_l$ driving the instability. In this exceptional case the amplitude
expansion is entirely free of singularities.
\begin{prop} If $f_{nl}=0$ for $n>1$, then the coefficients $p_j$ in
{\rm (\ref{eq:pj})} are all nonsingular:\label{prop:kur}
\begin{equation}
\lim_{\gamma\rightarrow0^+}\,|p_j|<\infty
\end{equation}
for $j\geq0$.
\end{prop}
\noindent  {\em {\bf Proof.}}\begin{quote} The functions $I_{m,j}(\omega)$ and
$h_{m,j}(\omega)$ have poles in the upper half plane only and consequently the
integrals appearing in the
coefficients $p_j$ are all free of pinching singularities in the limit
$\gamma\rightarrow0^+$. Also $\Gamma_{m,j}$ is nonsingular from
(\ref{eq:identity1}) in light of the discussion following
(\ref{eq:identity10}). Hence there are no singularities.

{\bf $\Box$}\end{quote}

The absence of singularities results immediately in the conclusion $\beta=1/2$
for the scaling exponent.

\section{Evaluation of Daido's order function}\label{sec:daido}
Daido's definition of the order function $H(\theta)$ is a direct generalization
of the familiar order parameter of Kuramoto and is motivated by the mean field
form of the phase dynamics (\ref{eq:gcoupled}).\cite{dainew,dainew96,dai2}  He
assumes there is a component of the population collectively entrained with
frequency $\Omega_e$ and introduces the shifted phases
$\psi_j\equiv\theta_j-\Omega_e t$, then phase equations
(\ref{eq:gcoupled}) can be rewritten
\begin{equation}
\dot{\psi}_i=\Delta_i-H({\psi}_i,t) +
\xi_i(t)\label{eq:gcoupledmf}
\end{equation}
where $\Delta_i\equiv\omega_i-\Omega_e$ and $H(\theta,t)$ is the mean field
\begin{equation}
H(\theta,t)\equiv -\sum_{n=-\infty}^{\infty}\,f_n\,e^{-in\theta}
\left[\frac{1}{N}\sum_{j=1}^{N}\,e^{in(\theta_j(t)-\Omega_e t)}\right].
\label{eq:mf}
\end{equation}
The additive noise term $\xi_i(t)$ in (\ref{eq:gcoupled})is omitted in the
analysis, and for large $N$, Daido assumes $\Omega_e$ can be chosen so that the
limit
\begin{equation}
H(\theta)\equiv \lim_{t\rightarrow\infty}H(\theta,t)\label{eq:daido}
\end{equation}
exists. The time-asymptotic mean field then defines the order function
$H(\theta)$. A norm of $H$,
\begin{equation}
\|H\|^2=\int_0^{2\pi}d\theta\,H(\theta)^2/2\pi,
\end{equation}
serves as an order parameter for models with multi-component couplings.

It is interesting to evaluate $H$ using our continuum description. The average
over the population in
(\ref{eq:mf}) can
be expressed in terms of $\rho(\theta,\omega,t)$:
\begin{equation}
\lim_{N\rightarrow\infty}
\frac{1}{N}\sum_{j=1}^{N}\,e^{in(\theta_j(t)-\Omega_e t)}
=\int_0^{2\pi}\,d\theta\int_{-\infty}^{\infty}\,d\omega\,g(\omega)
\rho(\theta,\omega,t) e^{in(\theta-\Omega_e t)},
\end{equation}
and this yields a general expression for the order function
\begin{equation}
H(\theta)\equiv -\lim_{t\rightarrow\infty}
\sum_{n=-\infty}^{\infty}\,f_n\,e^{-in\theta}
\int_0^{2\pi}\,d\theta'\int_{-\infty}^{\infty}\,d\omega'\,g(\omega')
\rho(\theta',\omega',t) e^{in(\theta'-\Omega_e t)}.\label{eq:ofcn}
\end{equation}
Let $H^u$ denote the order function obtained from solutions on the unstable
manifold (\ref{eq:umfdb}); a straightforward substitution from (\ref{eq:umfdb})
then yields
\begin{eqnarray}
H^u(\theta)/2\pi&=& -\lim_{t\rightarrow\infty}
\left\{f_l\,e^{-il(\theta+\Omega_et)}\left[\alpha(t)^\ast+
\alpha(t)^\ast\sigma(t)\int_{-\infty}^{\infty}\,d\omega \,g(\omega)
h_1(\omega,\sigma(t))^\ast\right]\right.\label{eq:ofcnum}\\
&&\hspace{0.25in}\left.
+\sum_{m=2}^\infty\,f_{ml}\,e^{-iml(\theta+\Omega_et)}
(\alpha(t)^\ast)^m\int_{-\infty}^{\infty}\,d\omega \,g(\omega)
h_m(\omega,\sigma(t))^\ast)+{\rm cc}\right\};\nonumber
\end{eqnarray}
note that we have set $f_0=0$ for simplicity.

Daido's assumption of time-independent
behavior (\ref{eq:daido}) is naturally satisfied when the system is
described by a ``rotating wave'' solution of the form (cf. (\ref{eq:rwsoln}))
\begin{equation}
\alpha_{rw}(t)=\gamma^\beta\,R_{rw}\,e^{-i\phi_0}\,e^{-i\Omega_{rw}t}
\label{eq:rw}
\end{equation}
with $R_{rw}$ and $\Omega_{rw}$ the time-independent amplitude and frequency,
respectively. For these asymptotic states by choosing $\Omega_e=\Omega_{rw}/l$,
the limit in (\ref{eq:ofcnum}) exists and the order function is
\begin{equation}
\frac{H^u_{rw}}{2\pi}= \left[
f_l^\ast\,\gamma^\beta
R_{rw}\,e^{-i\phi_0}\,(1+\gamma^{2\beta}R_{rw}^2\Gamma_1)
e^{il\theta}+\sum_{m=2}^\infty\,
f_{ml}^\ast\,\gamma^{m\beta}R_{rw}^{m}\,e^{-im\phi_0}
\,\Gamma_m\,e^{iml\theta}\right]
+{\rm cc}
\end{equation}
with norm given by
\begin{equation}
\|H\|^2=(2\pi)^2\gamma^{2\beta}R_{rw}^2\,\left[|f_l|^2\,
(1+\gamma^{2\beta}R_{rw}^2\Gamma_1)^2
+\sum_{m=2}^\infty\,|f_{ml}|^2\,
\gamma^{(2m-2)\beta}R_{rw}^{2m-2}\,\Gamma_m^2\right].\label{eq:rwnorm}
\end{equation}

Our results on the possible values of $\beta$ imply corresponding predictions
for the scaling of $\|H\|$ near onset. There are several different cases
depending on the coupling and whether or not noise is included;
a summary appears in Table II.

\begin{enumerate}

\item Suppose that  either  $D>0$, or  $D=0$ and we have a ``single component''
coupling: $f_l\neq0$  and $f_{ml}=0$ for $m>1$. Then
as $\gamma\rightarrow0^+$, there are no singularities  and (\ref{eq:rwnorm})
gives
\begin{equation}
\|H\|=2\pi\gamma^{\beta}R_{rw}\,|f_l|\,\left[1+{\cal O}(\gamma^{2\beta})\right]
\sim(K-K_c)^\beta\label{eq:rwnormkur}
\end{equation}
with $\beta=1/2$.
In each of these
circumstances, the prediction $\|H\|\sim(K-K_c)^{1/2}$ is in agreement with
Daido's analysis of $\|H\|$.\cite{dainew96}

\item Suppose that $D=0$ and we have the general case with
$f_l\neq0$  and $f_{2l}\neq0$. The coefficients of the amplitude expansion have
the generic singularities
(\ref{eq:pjasy}) and the coefficients $\Gamma_m$ are also singular
(\ref{eq:ghat}). Substituting (\ref{eq:ghat}) into (\ref{eq:rwnorm})
we find
\begin{equation}
\|H\|^2=(2\pi)^2\gamma^{2\beta}R_{rw}^2\,\left[|f_l|^2\,
(1+\gamma^{2(\beta-1)}R_{rw}^2\hat{\Gamma}_1(R_{rw}^2))^2
+\sum_{m=2}^\infty\,|f_{ml}|^2\,\gamma^{2(m-1)(\beta-1)}
R_{rw}^{2m-2}\,\hat{\Gamma}_m^2\right];\label{eq:rwnormsing}
\end{equation}
in this case we have determined that $\beta=1$ which yields
\begin{equation}
\|H\|=2\pi\gamma\, R_{rw}\,
\left[|f_l|^2\,(1+R_{rw}^2\hat{\Gamma}_1)^2
+\sum_{m=2}^\infty\,|f_{ml}|^2\,R_{rw}^{2m-2}\,\hat{\Gamma}_m^2\right]^{1/2}
\sim(K-K_c)^1.\label{eq:rwnormfinal}
\end{equation}
This is the scaling for $\|H\|$ Daido found for transitions at $l=1$ when
there was no noise and the coupling had components at $l=2$. We obtain this
result for a transition
at arbitrary $l$ if the coupling satisfies $f_{2l}\neq0$ and $D=0$.

\item Suppose that $D=0$ and we have the special case with
$f_l\neq0$, $f_{2l}=0$,  and $f_{3l}\neq0$.  In this case, our analysis of the
recursion relations leaves considerable uncertainty in the value of $\beta$ as
indicated by the range (\ref{eq:range}).

\item Suppose that $D=0$ and we have the special case with O(2) symmetry and
also a coupling that satisfies $f(\phi)=-f(\phi+\pi/l)$. This requires
$f_n=0$ unless $n$ is an odd multiple of $l$ and the non-zero components must
be imaginary. In this case again one has typically $f_l\neq0$ and
$f_{3l}\neq0$, and our analysis of the recursion relations specifies $\beta$
only
within the range (\ref{eq:range}). However this range  nevertheless {\em
excludes}
the value $\beta=1/2$ Daido obtains for such symmetric
systems.\cite{dainew,dai2,dai93b}
The reason for this apparent discrepancy has not been found.
\end{enumerate}

\section{Discussion}

Our analysis shares a key conclusion with Daido's study of scaling based
on the order function. In the absence of noise, the presence of a coupling
component $f_{2l}$ at the first harmonic of the critical mode number $l$ will
slow the onset of synchronized behavior so that the phase-locked component
of the population scales like $(K-K_c)$ rather than $(K-K_c)^{1/2}$ as in
the single-component coupling models.
However, both approaches suffer a common deficiency;  neither provides a
simple explanation of {\em why}
the onset of synchronization is slower if the coupling has a second harmonic
component ($f_{2l}\neq0$). This shift seems to be a true
many-body effect requiring large $N$, but a dynamical mechanism has not been
identified.  Perhaps further numerical studies will shed some light on this
issue.

For a general coupling with $f_{2l}\neq0$, our calculations show two effects of
adding noise to the phase dynamics. The diffusion term tends to suppress the
linear instability so that $K_c$ increases, but in addition the noise
prevents the poles from actually reaching the real axis and therefore removes
the coefficient singularities; this modifies the scaling $(K-K_c)^\beta$ of the
synchronized state  changing $\beta$
from $1$ to $1/2$. The interaction of these two effects raises the
intriguing possiblity that under some circumstances, adding noise could
actually yield an enhancement in the level of synchronization. In some recent
studies
of mean field oscillator models it has been noted that noise can enchance
a collective oscillation.\cite{hr94}

The linear stability analysis of section \ref{sec:linear} clearly shows that
for multi-component couplings one can expect multiple linear instabilities
and this will allow for studies of the interaction between two
different synchronizing transitions. In other settings the study of
simultaneous linear instabilities, or codimension-two mode interactions,
has been one of the keys to obtaining an analytical theory
capable of describing the system beyond the threshold of the first instability.
Such a theory can identify possibilities of secondary bifurcation and predict
the appearance of states with nontrivial time-dependence.
Despite the widespread interest in synchronization, and the large
literature on the Kuramoto model, there has been no analysis of the
codimension-two bifurcations that arise when more than one mode of
synchronization is possible.

Although the justification for a phase model varies from one context to
another,
in the setting of coupled oscillators it requires two essential
assumptions: weak coupling and a frequency spread $g(\omega)$ that is
``sufficiently'' small.\cite{erkop,kur,ashwin}
Moving beyond these assumptions necessitates
restoring the dynamics of the oscillator amplitudes $r_i(t)$, and this
adds an additional term to the kinetic equation of the schematic form
\begin{equation}
\frac{\partial}{\partial r}\cdot\left(\dot{r}\rho\right)
\end{equation}
where $r=(r_1,\ldots,r_N)$ are the oscillator amplitudes and $\dot{r}$ is
the non-stochastic component of the amplitude dynamics. In this situation,
$\rho$ describes the joint distribution of oscillator phases and amplitudes.
The effect of such a term on the synchronizing transition can be studied by
analyzing the effect on the singularities of the nonlinear coefficients.

\section{\hspace{0.125in}Acknowledgements}
This work supported by NSF grant PHY-9423583.

\appendix
\section{Proof of Proposition \ref{prop:sing}}\label{app:proof}
Let $z_0=(\Omega+i\gamma)/l$ denote the root in (\ref{eq:poles1})  -
(\ref{eq:poles2}). If $mn=0$ then there is no pinching singularity and the
integral in (\ref{eq:sing0}) has a finite limit; hence the limit is zero if
$J>0$.

Assume $mn>0$, then for integrals with $m+n=2$ the limit can be evaluated from
the Plemej formula\cite{musk} after a partial fraction
expansion. The expansion isolates the singularity at $\gamma=0$,
\begin{equation}
\int^\infty_{-\infty}\,\frac{d\omega\,\phi(\omega)}{(\omega-\alpha_1)
(\omega-\beta_1^\ast)}=\frac{-i/\gamma}{\delta_1+\zeta_1+2/l}
\left[\int^\infty_{-\infty}\,\frac{d\omega\,\phi(\omega)}{(\omega-\alpha_1)}-
\int^\infty_{-\infty}\,\frac{d\omega\,\phi(\omega)}{(\omega-\beta_1^\ast)}
\right],\label{eq:pfexp}
\end{equation}
and the limit (\ref{eq:sing0}) yields
\begin{eqnarray}
\lim_{\gamma\rightarrow0^+}\gamma
\left|\int^\infty_{-\infty}\,\frac{d\omega\,\phi(\omega)}{(\omega-\alpha_1)
(\omega-\beta_1^\ast)}\right|&=&
\frac{1}{(\delta_1+\zeta_1+2/l)}\lim_{\gamma\rightarrow0^+}\left|
\int^\infty_{-\infty}\,\frac{d\omega\,\phi(\omega)}{(\omega-\alpha_1)}-
\int^\infty_{-\infty}\,\frac{d\omega\,\phi(\omega)}{(\omega-\beta_1^\ast)}
\right|\nonumber\\
&&\nonumber\\
&=&\frac{2\pi|\phi(\Omega/l-Kf_0)|}{(\delta_1+\zeta_1+2/l)}.
\end{eqnarray}

The partial fraction expansion in (\ref{eq:pfexp}) expresses an integrand of
index 1 in
terms of integrands of index 0. More generally, for integrands
\begin{equation}
\int^\infty_{-\infty}\,d\omega\,\phi(\omega)\;{\cal G}(\omega)
\end{equation}
with  index $m+n-1>1$, by expanding the integrand in partial fractions, they
can be re-expressed as a sum of two integrands with index $m+n-2$
that are multiplied by an overall factor of $\gamma^{-1}$. Thus, if the partial
fraction integrands of index $m+n-2$
satisfy  (\ref{eq:sing0}) with $J=m+n-2$, then the original integrand will
satisfy (\ref{eq:sing0}) with $J=m+n-1$; the value of $J$ must be incremented
by one to allow for the factor of $\gamma^{-1}$.

A simple induction
argument along these lines establishes (\ref{eq:sing0}) for all $m+n\geq1$.  In
the limit in (\ref{eq:sing0}) is found to be non-zero in general so, barring an
accidental cancellation, if $J$ is replaced by $J-1$ then the modified limit
will diverge as $\gamma^{-1}$.

\section{Derivation of the model}\label{app:deriv}
Basically, the kinetic equation arises as the
first member of a coupled hierarchy of equations quite similar to the
BBGKY hierarchy well known in the kinetic theory of gases and this
is probably the most elementary systematic approach. There are
alternative derivations using path integral methods.\cite{bonilla}

The solution of the system of $N$ stochastic differential equations in
(\ref{eq:gcoupled}) is a Markov process with transition probability ${\cal
P}({\bf\theta},{\bf\omega},t|{\bf\theta}',{\bf\omega}',t')$ where
${\bf\theta}=(\theta_1,\ldots,\theta_N)$ and
${\bf\omega}=(\omega_1,\ldots,\omega_N)$.\cite{doer} This transition
probability
satisfies a linear Fokker-Planck equation
\begin{equation}
\frac{\partial {\cal P}}{\partial t}+\frac{\partial}{\partial {\bf\theta}}\cdot
\left(\dot{{\bf\theta}}{\cal P}\right)=
D\frac{\partial}{\partial
{\bf\theta}}\cdot\frac{\partial{\cal P}}{\partial {\bf\theta}}\label{eq:fp}
\end{equation}
subject to the initial condition
\begin{equation}
{\cal P}({\bf\theta},{\bf\omega},t'|{\bf\theta}',{\bf\omega}',t')
=\delta^N({\bf\theta}-{\bf\theta}')\,\delta^N({\bf\omega}-{\bf\omega}').
\end{equation}
The notation $\dot{{\bf\theta}}=(\dot{\theta_1},\ldots,\dot{\theta}_N)$ in
(\ref{eq:fp}) refers only to the non-stochastic part of the dynamics
\begin{equation}
\dot{\theta}_i=\omega_i+\frac{K}{N}\sum^{N}_{j=1} f(\theta_j-\theta_i).
\end{equation}

Given an initial normalized ensemble of populations ${\cal
E}_0({\bf\theta},{\bf\omega})$, the ensemble for $t>t_0$ is determined
by
\begin{equation}
{\cal E}({\bf\theta},{\bf\omega},t)=\int\, d{\bf\omega}'\;\int
\,d{\bf\theta}'\;
{\cal P}({\bf\theta},{\bf\omega},t|{\bf\theta}',{\bf\omega}',t_0)\;
{\cal E}_0({\bf\theta}',{\bf\omega}');
\end{equation}
equivalently, this is can be written as an evolution equation for ${\cal
E}({\bf\theta},{\bf\omega},t)$
\begin{equation}
\frac{\partial {\cal E}}{\partial t}+\frac{\partial}{\partial {\bf\theta}}\cdot
\left(\dot{{\bf\theta}}{\cal E}\right)=
D\frac{\partial}{\partial
{\bf\theta}}\cdot\frac{\partial{\cal E}}{\partial {\bf\theta}}\label{eq:fp2}
\end{equation}
with initial condition ${\cal E}_0$. The original phase dynamics
(\ref{eq:gcoupled})
is unchanged if we exchange a pair of oscillators $(\theta_i,\omega_i)
\leftrightarrow(\theta_j,\omega_j)$ and it is natural to select initial
ensembles ${\cal E}_0$ with this same invariance under pairwise exchange. In
addition,
for our problem, the values of the frequencies $\omega_i$ are assumed to
obey a fixed distribution $g(\omega)$. This requirement necessitates a suitable
choice of initial ensemble:
\begin{equation}
{\cal E}_0({\bf\theta},{\bf\omega})=
G_{0}({\bf\theta},{\bf\omega})\prod_{j=1}^{N} g(\omega_j)
\label{eq:ens}
\end{equation}
where the normalization of ${\cal E}_0({\bf\theta},{\bf\omega})$ implies
\begin{equation}
1=\int\, d{\bf\theta}\;G_{0}({\bf\theta},{\bf\omega}).
\end{equation}

A reduced distribution $\rho_s$ for the ensemble is defined in the usual way
\begin{equation}
\rho_s(\theta_1,\ldots,\theta_s,\omega_1,\ldots,\omega_s,t)\equiv
\int\,d\theta_{s+1}\cdots\,d\theta_{N}\int\,d\omega_{s+1}\cdots\,d\omega_{N}\,
{\cal E}({\bf\theta},{\bf\omega},t),\label{eq:rdist}
\end{equation}
and one can derive a hierarchy of coupled equations expressing the evolution
of $\rho_s$ in terms of $\rho_s$ and $\rho_{s+1}$
by performing the integration in (\ref{eq:rdist}) on the evolution equation
(\ref{eq:fp2}). For our purposes it is
sufficient to find the equation for $\rho_1$; integrating (\ref{eq:fp2})
over $(\theta_2,\ldots,\theta_N)$ and $(\omega_2,\ldots,\omega_N)$ yields
\begin{eqnarray}
\lefteqn{\frac{\partial \rho_1}{\partial t}
+\frac{\partial}{\partial\theta_1}
\left[\rule{0in}{0.25in}(\omega_1+{Kf(0)}/{N})\rho_1
+\frac{K(N-1)}{N}\int d\theta_2\int d\omega_2 f(\theta_2-\theta_1)
\rho_2(\theta_1,\theta_2,\omega_1,\omega_2,t)\right]}
\hspace{0.5in}\nonumber\\
&&\hspace{4.25in}=D\frac{\partial^2 \rho_1 }{\partial \theta_1^2}.
\end{eqnarray}
In terms of the two-oscillator correlation function,
$$C(\theta_1,\theta_2,\omega_1,\omega_2,t)\equiv
\rho_1(\theta_1,\omega_1,t)\,\rho_1(\theta_2,\omega_2,t)
-\rho_2(\theta_1,\theta_2,\omega_1,\omega_2,t),$$
this becomes
\begin{eqnarray}
\lefteqn{\frac{\partial \rho_1}{\partial t}
+\frac{\partial}{\partial\theta_1}
\left[\rule{0in}{0.25in}(\omega_1+{Kf(0)}/{N})\rho_1
+\frac{K(N-1)\rho_1}{N}
\int d\theta_2\int d\omega_2 f(\theta_2-\theta_1)
\,\rho_1(\theta_2,\omega_2,t)\right]=}\hspace{0.25in}\nonumber\\
&&\hspace{0.25in}D\frac{\partial^2 \rho_1 }{\partial
\theta_1^2}+\frac{\partial}{\partial\theta_1}
\left[\rule{0in}{0.25in}\frac{K(N-1)}{N}\int d\theta_2\int d\omega_2
f(\theta_2-\theta_1)
C(\theta_1,\theta_2,\omega_1,\omega_2,t)\right].
\end{eqnarray}

We obtain an autonomous equation for $\rho_1$ by letting $N\rightarrow\infty$
and discarding the correlation term
\begin{eqnarray}
\frac{\partial \rho_1}{\partial t}
+\frac{\partial}{\partial\theta_1}
\left[\rule{0in}{0.25in}\omega_1\,\rho_1
+{K\,\rho_1}\,
\int d\theta_2\int d\omega_2 f(\theta_2-\theta_1)
\,\rho_1(\theta_2,\omega_2,t)\right]&=&D\frac{\partial^2 \rho_1 }{\partial
\theta_1^2}.\label{eq:kinetic}
\end{eqnarray}
In some cases, one can rigorously prove that with a mean field interaction
the effects of correlations are negligible in the limit
$N\rightarrow\infty$.\cite{zwan, daw,bonilla}) Whether this kind of result
holds
for the models considered here appears to be an unsettled issue.  Finally,
the ensemble choice in (\ref{eq:ens}) implies that
\begin{equation}
\int_0^{2\pi}\,d\theta_1\,\rho_1(\theta_1,\omega_1,t_0)=g(\omega_1),
\end{equation}
so it is natural to define $\rho(\theta_1,\omega_1,t)$ by
\begin{equation}
\rho_1(\theta_1,\omega_1,t)\equiv g(\omega_1)
\rho(\theta_1,\omega_1,t).
\end{equation}
Substituting this expression for $\rho_1$ into (\ref{eq:kinetic})
yields the kinetic equation (\ref{eq:eveqn}) - (\ref{eq:vel}).

It is interesting to note that
the choice of initial ensemble (\ref{eq:ens}) can be adapted to treat other
forms of quenched randomness.\cite{bonilla2} For example, we can allow for
random variation in the coupling strength or even in the sign of the coupling
between different pairs of phases.


\clearpage


\begin{table}
\caption{Order of calculation of $h_{m,j}(\omega)$ and $p_j$ from
$\psi(\omega)$. The flow of calculation of the $h_{m,j}(\omega)$ is indicated
by moving downward; the calculation of the $p_j$ proceeds by moving downward
and to the left. For example,  from $\psi(\omega)$, $h_{2,0}$ can be calculated
and then $p_1$ determined; $h_{1,0}$ and $h_{3,0}$ are calculated next from
$h_{2,0}$ and then $h_{2,1}$ can be evaluated from $\{h_{1,0}, h_{3,0}, p_1\}$.
This then determines $p_2$, and so forth. For $N\geq2$, $p_N$ requires prior
calculation of $h_{m,j}$ for $1\leq m\leq N+1$ and $0\leq j\leq N-m+1-2
\delta_{m,1}.$}
\vspace{0.25in}
\begin{tabular}{l|ccccccc}
   &$m=1$&$m=2$&$m=3$&$m=4$&$m=5$&$m=6$&$\cdots$\\ \hline
\\
$p_0$& $\psi(\omega)$   &   & & & & & \\
\hline
$p_1$&  -         &$h_{2,0}$& & & & &\\
\hline
     &$h_{1,0}$&            &$h_{3,0}$& & & &\\
$p_2$&            &$h_{2,1}$&            & & & & \\
\hline
     &$h_{1,1}$&            &            &$h_{4,0}$& & & \\
$p_3$&            &$h_{2,2}$&$h_{3,1}$&            & & &\\
\hline
     &$h_{1,2}$&            &         &      &$h_{5,0}$& &\\
$p_4$&            &$h_{2,3}$& $h_{3,2}$&$h_{4,1}$ && &\\
\hline
     &$h_{1,3}$&            &         &      &&$h_{6,0}$ &\\
$p_5$&            &$h_{2,4}$& $h_{3,3}$&$h_{4,2}$ &$h_{5,1}$& &\\
\hline
     &&            &         &      && &\\
$\vdots$&&& & && &\\
\end{tabular}
\label{table1}
\end{table}

\begin{table}
\caption{Comparison of the exponent $\beta$ obtain from the amplitude
equation with the exponent $\beta_D$ found from Daido's order function.
We assume the synchronizing transition occurs at mode number $l$ for
a coupling with $f_l\neq0$. The bifurcation has O(2) symmetry when the coupling
and frequency distribution satisfy $f(\phi)=-f(-\phi)$ and
$g(\omega)=g(-\omega)$, respectively.}
\vspace{0.25in}
\begin{tabular}{lccccc}
Noise&Symmetry&Coupling&$\beta$&$\beta_D $ (for $l=1$)&Comment\\ \hline
\\
$D>0$&SO(2) or O(2)& arbitrary   &  1/2 & - & generic case with noise\\ \hline
\\
$D=0$&SO(2) or O(2)& $f_{2l}\neq0$ &  1 & 1 & generic case without noise\\
\hline
\\
$D=0$&SO(2) or O(2)& $f_{2l}=0$, $f_{3l}\neq0$   &  [3/4,1] & - &\\ \hline
\\
$D=0$&O(2)& $f(\phi)=-f(\phi+\pi/l)$   &  [3/4,1] & 1/2& $\beta$ and $\beta_D$
differ \\ \hline
\\
$D=0$&SO(2) or O(2)& $f_{ml}=0,\,m>1$   &  1/2 & 1/2 & single component case\\
\hline
\\
\end{tabular}
\label{table2}
\end{table}



\newpage
\begin{thebibliography}{99}

\bibitem{erkop} G.B. Ermentrout and N. Kopell, Frequency plateaus in a chain of
weakly coupled oscillators, I, {\em SIAM J. Math Anal.} {\bf 15} (1984) 215.

\bibitem{kur} Y. Kuramoto, {\bf Chemical Oscillations, Waves, and Turbulence},
 Springer-Verlag, New York (1984).

\bibitem{ashwin} P. Ashwin and J.W. Swift, The dynamics of $n$ weakly coupled
identical oscillators, {\em J. Nonlinear Sci.} {\bf 2} 69.

\bibitem{win} A.T. Winfree, Biological rhythms and the behavior of populations
of coupled oscillators, { J. Theor. Biol.} {\bf 16} (1967) 15-42.

\bibitem{win2} A.T. Winfree, {\bf Geometry of Biological Time},
Springer-Verlag, New York (1990).

\bibitem{stro1} S.H. Strogatz, Norbert Wiener's Brain Waves,
in Lecture Notes in Biomathematics, Vol. 100, ed. S. Levin,
(Springer Verlag, New York, 1994).

\bibitem{gray} C.M. Gray, P. Konig, A.K. Engel and W. Singer, Oscillatory
responses in cat visual cortex exhibit inter-columnar synchronization which
reflects global stimulus properties, {Nature} {\bf 338} (1989) 334-337.

\bibitem{wcs} K. Wiesenfeld, P. Colet, S.H. Strogatz, Synchronization
transitions in a disordered Josephson series array,  {Phys. Rev. Lett.} {\bf
76} (1996) 404.

\bibitem{ssw} J.W. Swift, S.H. Strogatz, and K. Wiesenfeld,
Averaging of globally coupled
oscillators,  {Phys. Rev. E} {\bf 51} (1995) 1020.

\bibitem{ws} K. Wiesenfeld and J.W. Swift, Averaged equations for Josephson
junction series arrays,  {Physica D} {\bf 55} (1992) 239.

\bibitem{kn} Y. Kuramoto and I. Nishikawa, Statistical macrodynamics of large
dynamical systems. Case of a phase transition in oscillator communities, {\em
J. Stat. Phy.} {\bf 49} (1987) 569-605.

\bibitem{hmm} D Hansel, G. Mato, and C. Meunier, Phase dynamics for weakly
coupled Hodgkin-Huxley neurons, { Europhysics. Lett.} {\bf 23} (1993) 367-372.

\bibitem{dainew} H. Daido, Generic scaling at the onset of macroscopic mutual
entrainment in limit-cycle oscillators with uniform all-to-all coupling, {
Phys. Rev. Lett.} {\bf 73} (1994) 760-763.

\bibitem{jdc3} J.D. Crawford, Scaling and singularities in the entrainment of
globally coupled oscillators,  { Phys. Rev. Lett.} {\bf 74}
(1995) 4341-4344.

\bibitem{dainew96} H. Daido, Onset of cooperative entrainment in limit-cycle
oscillators with uniform all-to-all interactions: bifurcation of the order
function, {Physica D} {\bf 91} (1996) 24-66.

\bibitem{sak} H. Sakaguchi, Cooperative phenomena in coupled oscillator systems
under external fields, { Prog. Theor. Phys.} {\bf 79} (1988) 39-46.

\bibitem{sm} S.H. Strogatz and R. Mirollo, Stability of incoherence in a
population of coupled oscillators, { J. Stat. Phys.} {\bf 63} (1991) 613-635.

\bibitem{smm} S.H. Strogatz, R. Mirollo and P.C. Matthews, Coupled nonlinear
oscillators below the synchronization threshold: relaxation by generalized
Landau damping, { Phys. Rev. Lett.} {\bf 68} (1992) 2730-2733.

\bibitem{jdc0} J.D. Crawford, Amplitude expansions for instabilities in
populations of globally-coupled oscillators, { J. Stat. Phys} {\bf 74} (1994)
1047-1084.

\bibitem{craw5} J.D. Crawford, Introduction to bifurcation theory,
{\em Rev. Mod. Phys.} {\bf 63} (1991) 991-1037.

\bibitem{iv}  A. Vanderbauwhede and G. Iooss, 1992, Centre manifold theory in
infinite dimensions, {\bf Dynamics Reported}, Vol. 1, Springer-Verlag, New
York, 125-163.

\bibitem{jdc1} J.D. Crawford, Universal trapping scaling on the unstable
manifold of a collisionless electrostatic mode,  { Phys. Rev. Lett.} {\bf 73}
(1994) 656-659.

\bibitem{jdc2} J.D. Crawford, Amplitude equations for electrostatic waves:
universal singular behavior in the limit of weak
instability, { Phys. Plasmas} {\bf 2} 97-128 (1995).

\bibitem{jdcaj} J.D. Crawford and A. Jayaraman, Nonlinear saturation of
electrostatic waves: mobile ions modify
trapping scaling, { Phys. Rev. Lett.} {\bf 77}
 3549 (1996).

\bibitem{case2} K. Case, Stability of inviscid plane Couette flow, { Phys. Fl.}
{\bf 3} 143 (1960).

\bibitem{briggs} R.J. Briggs, J.D. Daugherty, and R.H. Levy, Role of Landau
damping in crossed-field electron beams and inviscid shear flow, { Phys. Fl.}
{\bf 13}  421-432 (1970).

\bibitem{pegowein1} R. Pego and M.I. Weinstein, { Phil. Trans. R. Soc. Lond.
A} {\bf 340} (1992) 47-94.

\bibitem{pegowein2} R. Pego and M.I. Weinstein,  in {\em Differential Equations
with Applications to Mathematical Physics}, W.F. Ames, E.M. Harrell II, and
J.V. Herod, eds., Academic Press, Orlando, 1993. pp. 273-286.

\bibitem{pegwein3} R. Pego, P. Smereka, and M.I. Weinstein, Oscillatory
instability of solitary waves in a continuum model of lattice vibrations,
{Nonlinearity}, {\bf 8} 921-941 (1995).

\bibitem{russo} G. Russo and P. Smereka, Kinetic theory for bubbly flow I:
collisionless case, {SIAM J. Appl. Math.}, {\bf 56} (1996).

\bibitem{dai93}  H. Daido, Critical conditions of macroscopic mutual
entrainment in uniformly coupled limit-cycle oscillators,
{ Prog. Theor. Phys.} {\bf 89} (1993) 929-934.

\bibitem{ch} J.D. Crawford and P.D. Hislop, Application of the method of
spectral deformation to the Vlasov-Poisson system, {\em Ann. Phys.} {\bf 189}
(1989) 265-317.

\bibitem{gss} M. Golubitsky, I. Stewart, and D.G. Schaeffer, {\bf
Singularities and Groups in Bifurcation Theory}: Vol. II, Appl.
Math. Sci. {\bf 69}, Springer-Verlag, New York (1988).

\bibitem{dai2} H. Daido, Order function and macroscopic mutual entrainment in
uniformly coupled limit-cycle oscillators, { Prog. Theor. Phys.} {\bf 88}
(1992) 1213-1218.

\bibitem{dai93b} H. Daido, A solvable model of coupled limit-cycle oscillators
exhibiting partial perfect synchrony and novel frequency spectra,, {Physica D}
{\bf 69} (1993) 394-403.

\bibitem{hr94} V. Hakim and W.-J. Rappel, Noise-induced periodic behavior
in the globally coupled complex Ginzburg-Landau equation,
{\em Europhys. Lett.} {\bf 27} (1994) 637.

\bibitem{musk} N. I. Muskhelishvili,
{\bf Singular Integral Equations} (Noordhoff, Groningen,
 The Netherlands, 1953), pp. 56-61.

\bibitem{bonilla} L.L. Bonilla, Stable nonequilibrium probability
densities and phase transitions for mean-field
models in the thermodynamic limit,
{\em J. Stat. Phys.} {\bf 46} (1987) 659.

\bibitem{doer} C.R. Doering, Modelling complex systems: stochastic processes,
stochastic differential equations, and Fokker-Planck equations, in
{\bf Lectures in Complex Systems}, Eds.
L. Nadel and D. Stein, (Addison Wesley, New York, 1991), pp. 3-51.

\bibitem{zwan} Rashmi C. Desai and Robert Zwanzig, Statistical mechanics of a
nonlinear stochastic model, {\em J. Stat. Phy.} {\bf 19}
(1978) 1-24.

\bibitem{daw} D.A. Dawson,  Critical dynamics and fluctuations for a mean-field
model of cooperative behavior, {\em J. Stat. Phy.} {\bf 31}
(1983) 29.

\bibitem{bonilla2} L.L. Bonilla, Glassy synchronization in a population of
coupled oscillators,
{\em J. Stat. Phys.} {\bf 70} (1993) 921.

\end{thebibliography}
\end{document}